\documentclass[%
aps, reprint, a4paper, twocolumn, superscriptaddress, amsmath, amssymb, floatfix, nofootinbib
]{revtex4-2}

\usepackage{lmodern}
\usepackage{anyfontsize}
\usepackage{microtype} 
\usepackage{amsthm}
\usepackage{graphicx}
\usepackage{epstopdf}
\usepackage[table,dvipsnames]{xcolor}
\usepackage{braket}
\usepackage[colorlinks,bookmarksopen=false]{hyperref}
\usepackage[caption=false]{subfig}
\usepackage{bbm}
\usepackage{makecell}
\usepackage{multirow}
\usepackage{lipsum, blindtext}
\usepackage{xr}
\usepackage{MnSymbol}
\usepackage{siunitx}
\usepackage[normalem]{ulem}
\usepackage{cancel}
\usepackage{subdepth} 
\usepackage{orcidlink} 

\hypersetup{%
  pdfstartview={FitW},
  linkcolor=blue,
  citecolor=blue,
  filecolor=black,
  menucolor=black,
  urlcolor =blue
}

\newcommand{\im}{\ensuremath{\textup{i}}}

\newcommand{\op}[1]{\ensuremath{\mathsf{#1}}}

\newcommand{\uop}{\op{\mathbbm{1}}}

\newtheorem*{mydef*}{Definition}
\newtheorem*{myprop*}{Proposition}

\newtheorem*{mythm*}{Theorem}

\newcommand{\myvec}[1]{\boldsymbol{#1}}

\newcommand{\rmenc}{\mathrm{enc}}
\newcommand{\rment}{\mathrm{ent}}

\newcommand{\rmout}{\mathrm{out}}

\newcommand{\rmx}{\mathrm{x}}
\newcommand{\rmy}{\mathrm{y}}
\newcommand{\rmz}{\mathrm{z}}

\newcommand{\rmq}{\mathrm{q}}

\newcommand{\rminit}{\mathrm{init}}
\newcommand{\rmctrl}{\mathrm{ctrl}}
\newcommand{\rmtrgt}{\mathrm{trgt}}

\newcommand{\rmc}{\mathrm{c}}

\newcommand{\rmlayer}{\mathrm{layer}}

\newcommand{\rmtrain}{\mathrm{train}}
\newcommand{\rmop}{\mathrm{op}}
\newcommand{\rmrep}{\mathrm{rep}}
\newcommand{\rmhad}{\mathrm{had}}
\newcommand{\rmpqc}{\mathrm{pqc}}
\newcommand{\rmtest}{\mathrm{val}}
\newcommand{\rmparam}{\mathrm{param}}
\newcommand{\rmtrue}{\mathrm{true}}
\newcommand{\rmpred}{\mathrm{pred}}

\newcommand{\change}[1]{#1}
\newcommand{\xchange}[1]{}

\begin{document}

\title{%
  Quantum Neural Networks in Practice: A Comparative Study with Classical Models from Standard Data Sets to Industrial Images
}

\author{Daniel Basilewitsch\orcidlink{0000-0002-3935-8347}}
\affiliation{%
  Quantum Applications Group, TRUMPF SE + Co.\ KG, Johann-Maus-Straße 2, 71254
  Ditzingen, Germany
}

\author{Jo\~{a}o F. Bravo\orcidlink{0000-0002-8526-5010}}
\email{joao.bravo@iao.fraunhofer.de}
\affiliation{%
 Fraunhofer IAO, Nobelstraße 12, 70569 Stuttgart, Germany
}

\author{Christian Tutschku}
\affiliation{%
 Fraunhofer IAO, Nobelstraße 12, 70569 Stuttgart, Germany
}

\author{Frederick Struckmeier}
\affiliation{%
  Quantum Applications Group, TRUMPF SE + Co.\ KG, Johann-Maus-Straße 2, 71254
  Ditzingen, Germany
}

\date{\today}

\begin{abstract}
  In this study, we compare the performance of randomized classical and quantum neural networks (NNs) as well as classical and quantum-classical hybrid convolutional neural networks (CNNs) for the task of \change{supervised} binary image classification. \change{We keep the employed quantum circuits compatible with near-term quantum devices and} use two distinct methodologies: applying randomized NNs on dimensionality-reduced data, and applying CNNs to full image data. We evaluate these approaches on three \change{fully-classical} data sets of increasing complexity: an artificial hypercube data set, MNIST handwritten digits and \xchange{real-world} industrial images \change{of practical relevance. Our study's central goal is to shed more light on how quantum and classical models perform for various binary classification tasks and on what defines a good quantum model. To this end, our study involves a correlation analysis between classification accuracy and quantum model hyperparameters, and an analysis on the role of entanglement in quantum models, as well as on the impact of initial training parameters.} \xchange{We analyze correlations between classification accuracy and quantum model hyperparameters, including the number of trainable parameters, feature encoding methods, circuit layers, entangling gate type and structure, gate entangling power, and measurement operators. For random quantum NNs, we compare their performance against literature models.} We find classical and quantum-\change{classical} hybrid models achieve statistically-equivalent classification accuracies across most data sets with no approach \change{consistently outperforming the other. Interestingly, we observe that quantum NNs show lower variance with respect to initial training parameters and that the role of entanglement is nuanced. While incorporating entangling gates seems to be generally advantageous, we also observe that their (optimizable) entangling power is not correlated with model performance. We also observe an inverse proportionality between the number of entangling gates and the average gate entangling power.} \xchange{We observe that quantum models show lower variance with respect to initial training parameters, suggesting better training stability. Among the hyperparameters analyzed, only the number of trainable parameters showed a positive correlation with the model performance. Around 94$\%$ of the best-performing quantum NNs had entangling gates, although for hybrid CNNs, models without entanglement performed equally well but took longer to converge. Cross-data set performance analysis revealed limited transferability of quantum models between different classification tasks.} Our study provides an industry perspective on quantum machine learning for practical \change{binary} image classification tasks, highlighting both current limitations and potential avenues for further research in quantum circuit design, entanglement utilization, and model transferability across varied applications. \\[5mm]
\end{abstract}

\maketitle

\section{Introduction}

Quantum machine learning~\cite{Schuld2014, Biamonte2017} is one of the various
subfields of quantum computing promising advantages over its classical
counterpart, for instance due to proposals regarding more efficient data
encoding~\cite{Lloyd2013, PRL.113.130503}, improved model
expressivity~\cite{PRA.98.032309, PRA.101.032308, PRR.2.033125}, prediction
accuracy~\cite{Huang2021, PRL.126.190505, Jerbi2024} or by saving computational
resources~\cite{Liu2021}. These theoretical proposals have been
backed up by claims of experimental quantum advantage in learning
tasks~\cite{Riste2017, Huang2022}.
\change{%
While these works illustrate the various hopes associated with quantum machine learning, they often rely on very specific assumptions about the problem~\cite{Jerbi2024, Liu2021}, the availability of quantum data~\cite{PRL.126.190505, Riste2017, Huang2022, PhysRevLett.134.120803}, or require quantum devices beyond the near-term regime~\cite{Lloyd2013, PRL.113.130503}. While there has been
}
undeniable progress in the field of
quantum machine learning --- both on the algorithmic as well as on the hardware
side --- doubts about the field's focus on quantum advantage have been
previously raised~\cite{PRXQ.3.030101}. Ultimately, this culminated in a recent
work questioning the significance of quantum-machine-learning benchmark
studies~\cite{Bowles2024}, given their predominant use of overly simple data
sets paired with insufficiently-performing quantum hardware. Other
works~\cite{Cerezo2024, Bermejo2024, Angrisani2024} even suggest that much of
what was commonly believed to be classically hard or intractable to simulate
might be efficiently simulable on classical hardware after all. While this \xchange{may
require the field to reevaluate what ``quantum'' means for quantum machine
learning and how a future quantum computer may use ``quantumness''
advantageously for learning tasks,} \change{raises the question what ``quantum'' means for quantum machine learning,} both Refs.~\cite{Bowles2024, Cerezo2024}
independently call for more rigorous benchmarks employing less trivial data sets
in order to assess the potential of quantum machine learning.

In response to this critical need, our work directly addresses this challenge by employing not only standard benchmark data sets, but crucially incorporating \xchange{real-world} industrial data \change{of practical relevance} that represents \xchange{practical} machine-learning applications beyond academic test cases.
\change{%
To put our work into perspective, as pointed out earlier, many papers about (theoretical) quantum advantage require either very specific problem assumptions or quantum data. However, many machine-learning problems of practical relevance have neither very specific problem assumptions, nor quantum data. In the following, we therefore take an industrial perspective on quantum machine learning and want to investigate whether quantum models provide some kind of ``practical advantage'' for a supervised binary classification task on classical data, compared to classical models. We thereby measure ``practical advantage'' by the achievable classification accuracies and compare them between classical and quantum models. Since we believe that a fair comparison requires testing of various models and scenarios, we conceive our study as a larger benchmarking study that sheds light on the differences and properties of classical and quantum models and on what defines a good quantum model. We therefore deliver a robust benchmark study in the spirit of Refs.~\cite{Bowles2024, Cerezo2024}, using non-trivial data sets of practical relevance for the industry, while, at the same time, answering the question of how quantum machine learning models of small to near-term scale perform for such a task. We believe this study provides important insights towards auspicious future research directions.
}

\xchange{In this work} \change{In more detail}, we benchmark the performance of classical and quantum-machine-learning models in classifying three \change{classical} data sets of increasing data complexity.
The data sets include (i) an artificial data set based on linearly splitting
a (hyper-)cube in two, (ii) MNIST's handwritten digits~\cite{deng2012mnist} as
an easy but well-studied data set and (iii) images from laser cutting machines
as a challenging data set from the industry. Note that we use the first two data
sets merely as a means to benchmark and validate our employed methodologies,
since --- as mentioned in Refs.~\cite{Bowles2024, Bermejo2024} --- the potential
insights about the power and usefulness of quantum machine learning that can be
drawn from simple data sets are very limited. However, we still believe them to
be useful for validating the results we obtain for the practically-relevant
industrial data set on which we put the main focus when drawing conclusions on
quantum machine learning from an industry perspective. Moreover, since we are
also interested in testing the cross-data-set performance, assessed by the
classification accuracy, of some quantum models, we deem it beneficial to have
results for various data sets with different levels of complexity available.

In terms of methodologies, we employ both classical and quantum neural
networks~\cite{PRA.98.032309} as well as classical and quantum-classical hybrid
convolutional neural networks~\cite{Henderson2019} to classify any of the
mentioned data sets.
The dual-methodology approach used allows us to systematically analyze not just the relative performance of classical versus quantum models, but also to extract insights on which quantum architecture aspects contribute most to performance, how hyperparameters correlate with classification accuracy, and whether successful quantum models on one data set transfer well to others --- providing a comprehensive framework for quantum advantage assessment in realistic settings.
Our decision for parametrized quantum circuits is motivated
by their high expressivity~\cite{PRA.98.032309} which is amplified by techniques
like data re-uploading~\cite{PerezSalinas2020, PRA.103.032430} and trainable
data embeddings~\cite{Lloyd2020} while at the same time keeping quantum hardware
requirements moderate. This is in contrast to some early suggestions for quantum-machine-learning models based on amplitude encoding~\cite{PRL.113.130503,
Lloyd2013, Schuld2017}, which, despite being qubit-efficient, were later deemed
too resource-demanding to run on any near-term device due to their excessive
need of quantum gates for accurate state preparation~\cite{PRA.101.032308}. In
contrast to these advantages, parametrized quantum circuits suffer from barren
plateaus~\cite{McClean2018}, which prevents their efficient training at scale,
although much work has been put into mitigating or avoiding them, see e.g.\
Refs.~\cite{Grant2019, Cerezo2021NatComm, PRR.3.033090, Deshpande2024} to name
but a few. Some quantum models, such as quantum convolutional neural
networks~\cite{Cong2019}, provably avoid barren plateaus~\cite{PRX.11.041011}
but are classically efficiently simulable~\cite{Bermejo2024}. Unfortunately,
the latter are also rather resource-demanding to qualify for near-term
applicability and practical simulability within this study, which is why we
resort to quantum-classical hybrid convolutional neural
networks~\cite{Henderson2019} instead. Conceptually, parametrized quantum
circuits as well as quantum-classical hybrid convolutional neural networks
belong to the realm of variational quantum algorithms~\cite{Cerezo2021}, where
the execution of quantum circuits is carried out on actual quantum hardware,
while the calculation of iteratively-updated circuit parameters is performed on
classical computers. Due to these choices, our study reflects a current
perspective on what quantum-machine-learning methods, suitable for near-term
quantum devices, offer for classifying industry-level image data.

The remainder of this paper is structured as follows. In Sec.~\ref{sec:data} we
introduce the three different data sets in more detail, before
Sec.~\ref{sec:method} presents the main methodological framework based on random
neural networks and convolutional neural networks. The results based on the
former are shown in Sec.~\ref{sec:results:rnd}, while
Sec.~\ref{sec:results:conv} subsequently presents the results based on the
latter. Section~\ref{sec:conclusion} concludes the work by summarizing the key
findings, evaluating the implications of the results, and discussing the broader
significance of this type of benchmark.

\section{Data sets}
\label{sec:data}

In the following, we motivate and introduce the data sets used within this
study. These encompass two simple and well-studied data sets as well as
an \change{industrial} \xchange{real-world} image data set of \change{practical} \xchange{industrial} relevance. The associated task for all
data sets is a binary classification.

\begin{figure*}
  \centering
  \includegraphics{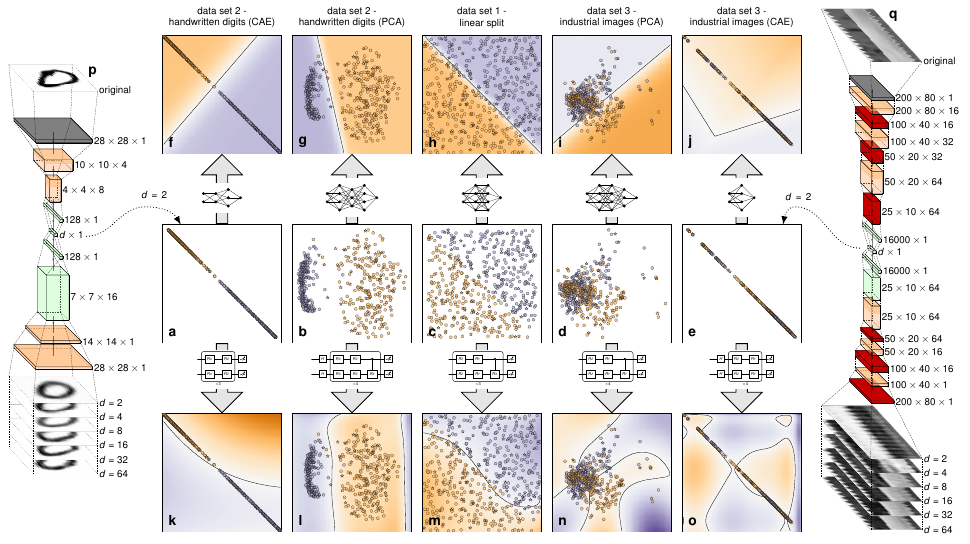}
  \caption{%
    Overview of data sets with $N=500$ and dimension $d = 2$ in panels (a)-(e),
    and the corresponding learned functions by classical and quantum neural
    networks in panels (f)-(j) and panels (k)-(o), respectively. The small
    neural networks in between panels (a)-(e) and (f)-(j) show the best
    performing architectures (from among the set of randomly generated neural
    networks). The small quantum neural networks in between panels (a)-(e) and
    (k)-(o) show the respective best performing quantum architectures (from
    among the set of randomly generated quantum neural networks). Data points
    represented by circles (stars) in panels (a)-(o) correspond to the training
    (validation) data set. In addition, panels (p) and (q) show the architecture
    for the convolutional autoencoder (CAE) used to reduce the data
    dimensionality for the images of the second and third data set,
    respectively, as well as a representative illustration of the original and
    reconstructed images for various latent dimensions $d$.
  }
  \label{fig:overview}
\end{figure*}

\subsection{Data set 1 --- linear split}
\label{subsec:data:linsplit}

The first data set is an artificial one and the only one not based on images. It
is chosen with the premise of simplicity in mind and such that the ideal
function, which classifies the data set perfectly, is known in advance. To this
end, we choose to consider a $d$-dimensional hypercube, defined by the set $[0,
1]^{d} \subseteq \mathbb{R}^{d}$, and linearly split it by the plane orthogonal
to the normal vector $\myvec{n} = (1, 1, \dots, 1)^{\top} / \sqrt{d} \in
\mathbb{R}^{d}$ which passes through the hypercube's center $(1, 1, \dots,
1)^{\top} / 2 \in \mathbb{R}^{d}$. In order to generate training and validation
data sets, we first place $N$ data points randomly within the hypercube such
that exactly $N/2$ points are located within each of the two sectors separated
by the hyperplane. These two sectors define the two labels that classify each
data point. Afterwards, we split the $N$ data points into training (validation)
data sets consisting of $N_{\rmtrain} = 0.8 N$ ($N_{\rmtest} = 0.2 N$) of the
total data points. We perform this splitting so that the number of data points
with each label within each subset is identical. In order to study the effect of
the hypercube's dimension $d$ and the data set size $N$, we generate
versions of the hypercube data set with $d \in \{2, 4, 8, 16, 32, 64\}$ and $N
\in \{200, 500, 1000, 2000\}$. An example with $d = 2$ and $N = 500$ is shown in
Fig.~\ref{fig:overview}~(c).

This hypercube data set has recently also been studied in a related work
exploring the performance of quantum-machine-learning methods~\cite{Bowles2024}.
In this work, the hypercube data turned out to be rather difficult to classify
by means of quantum-machine-learning methods, despite being a trivial problem for
classical machine learning. Hence, we use it in our study as an ideal means to
validate and benchmark our methodologies, introduced in Sec.~\ref{sec:method},
and to check whether it can reproduce those results qualitatively.

\subsection{Data set 2 --- handwritten digits}
\label{subsec:data:mnist}

As a second data set, we take the well-studied handwritten digits of the MNIST
database~\cite{deng2012mnist} as another means to test our methodologies.
However, in order to keep the classification problem binary, we restrict our
study to images of $0$'s and $1$'s. Similar to the hypercube's data sets
described in Sec.~\ref{subsec:data:linsplit}, we generate different versions of
the MNIST data set with $N \in \{200, 500, 1000, 2000\}$ total images taken
randomly from the $60000$ available images. We again ensure that each generated
data set has an equal number of $0$'s and $1$'s and that this ratio is
maintained once the total data set is split into training and validation sets
with $N_{\rmtrain} = 0.8 N$ and $N_{\rmtest} = 0.2 N$ data points, respectively.

However, while the hypercube's data sets are artificial and their data dimension
$d$ is freely choosable, each of the MNIST images are $28 \times 28$ by default
and hence consist of $784$ unique features. While a data dimensionality of $784$
is trivial to handle for classical neural networks, it is already challenging
for fully-quantum machine-learning approaches in the present day. This is
unfortunately true for its execution on actual quantum hardware as much as for
any quantum models' simulation on classical computers. Thus, to make the MNIST
images approachable for quantum neural networks, which are used in one of our
two methodologies employed to classify the data, we create
dimensionality-reduced versions of the data. To this end, we consider again two
different methods for dimensionality reduction as follows.

As a first method, we employ a principal component analysis (PCA). We fit it on
all available images of $0$'s and $1$'s in the MNIST database and only
afterwards apply it to the actual data sets used for training and validation
with a total number of $N \in \{200, 500, 1000, 2000\}$ images. Due to this, we
seek to prevent potential differences in the classification accuracy which
might emerge, e.g.\ due to differences in data set sizes, having their origin in
the PCA. We choose different levels of data reduction and fit PCAs using $d \in
\{2, 4, 8, 16, 32, 64\}$ components in order to explore its impact on
classification accuracies and to have identical dimensions to those generated
for the hypercube data. At last, note that while the original MNIST images are
rescaled and reshaped to vectors in $[0, 1]^{784}$ prior to applying the PCA,
this normalization is broken by the PCA and we manually renormalize the
PCA-reduced data vectors to $[0, 1]^{d}$ afterwards. This ensures that the final
classical or quantum neural networks carrying out the actual classification only
receive feature values from within $[0, 1]$. An example of the MNIST data that
has been dimensionality-reduced via PCA with $d = 2$ and $N = 500$ is shown in
Fig.~\ref{fig:overview}~(b).

As a second method for dimensionality reduction, we employ a convolutional
autoencoder (CAE). To this end, we design a classical convolutional encoder
architecture which reduces the $28 \times 28$ input images to a latent space of
dimension $d$ and, after that, we use a decoder architecture to reconstruct the
original image as best as possible. After having tested various architectures,
we find the architecture illustrated in Fig.~\ref{fig:overview}~(p) to work best
for the MNIST images. This has been judged on the visual similarity of the
original and reconstructed images, cf.\ the representative original and
reconstructed images shown in the top and bottom of Fig.~\ref{fig:overview}~(p),
respectively. Furthermore, note that similarly to the PCA, the training of the
CAE is first carried out using all available images of $0$'s and $1$'s from the
MNIST database and the trained CAE is only applied afterwards to the actual data
sets with a total number of $N \in \{200, 500, 1000, 2000\}$ data points. As
latent dimension, we again choose and train CAEs for $d \in \{2, 4, 8, 16, 32,
64\}$. An example of the MNIST data that has been reduced via a CAE with $d = 2$
and $N = 500$ is shown in Fig.~\ref{fig:overview}~(a). It is important to note
that the employed CAEs apply an additional normalization by enforcing the sum of
all features of the reduced $d$-dimensional data vector to be unity, thus
effectively further reducing its dimension by one.

\subsection{Data set 3 --- industrial images}
\label{subsec:data:trumpf}

\begin{figure}
  \includegraphics{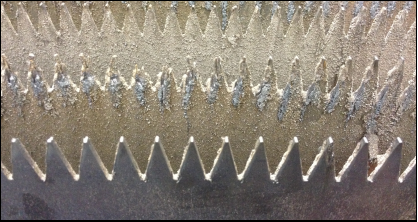}
  \caption{%
    Image of slats from a TRUMPF flatbed laser cutting machine. The slat in the
    front is in a good condition whereas the second slat has clear signs of wear
    and tear, such as tips that have melted away and slag splashes that adhere
    to the side. Image adapted from Ref.~\cite{Struckmeier2019}.
  }
  \label{fig:slats}
\end{figure}

Our third and most important, as well as most challenging, data set consists of
industrial images from laser cutting machines designed for cutting sheet metal.
Those machines feature supporting slats on which the sheet metal rests during
processing. Physically, these slats are thin metal strips that span the width of
the machine in a triangular pattern as illustrated in Fig.~\ref{fig:slats}. In
practice, their positions are regularly altered manually by adding or removing
them from their respective sockets. However, for machine operation, it is
advantageous to know their arrangement beforehand in order to take into account
possible side-effects that their arrangement may cause~\cite{Struckmeier2019},
thus necessitating measurements prior to operation. Given the prominence of
machine learning for automated image classification, visual methods have been
identified previously as the most promising approach for accurately detecting
the slats' positions~\cite{Struckmeier2020}. Hence, we consider it
a prototypical and industry-relevant image classification task.

In contrast to Fig.~\ref{fig:slats}, the slat images we use in the following are
already preprocessed and cut such as to focus on a single slat socket which may
or may not contain a slat. The classification problem is thus binary with two
difficulties arising due to varying camera angles from which the slat sockets
are being photographed as well as the different condition of the slats, cf.\
Fig.~\ref{fig:slats}. To ease data generation, TRUMPF has developed a model that
generates synthetic slat socket images given the CAD model of the entire laser
cutting machine and textures derived from real-world\change{-resembling} images using the Blender
rendering engine. This way, manual labeling effort of real-world images is
unnecessary and labeling errors are ruled out. Moreover, it allows to generate
data sets with large image numbers. Identically to the other data sets, we
consider different versions of this industry data set with $N \in \{200, 500,
1000, 2000\}$ total images, which are afterwards split into training and
validation data sets with $N_{\rmtrain} = 0.8 N$ and $N_{\rmtest} = 0.2 N$ data
points, respectively. For these two sets, we ensure again the same class
distribution as the original data set, as well as a similar number of images of
slat sockets photographed with different camera angles and conditions. All
images are gray-scale and normalized to $200 \times 80$ pixels independent of
the photo angle.

Similarly to the second data set, the images are too large to be directly
processable by fully-quantum models, such as a quantum neural network. Hence, we
create --- again similarly to the second data set --- dimensionality-reduced
versions with either a PCA or CAE as the reduction method and fit either of the
two methods to reduce the image data into $d$-dimensional data vectors. The
reduction is separately fitted for $d \in \{2, 4, 8, 16, 32, 64\}$ and happens
on all available slat images before being applied to the different versions of
the third data set. For the CAE, after trying various architectures, we find the
one illustrated in Fig.~\ref{fig:overview}~(q) to work best for the slat images.
An example of a $200 \times 80$ synthetic slat images is given at the top of
Fig.~\ref{fig:overview}~(q) and reconstructed versions from the various $d$'s at
the bottom.

\section{Methodology}
\label{sec:method}

As already indicated in Sec.~\ref{sec:data}, the images of the second and third
data sets are too large to be handled by fully-quantum machine-learning models.
Hence, we employ two different methodologies in the following to classify them.
The first of them being the usage of either classical and quantum neural
networks to classify the dimensionality-reduced version of the previously
introduced data sets. The other methodology being the usage of either classical
or quantum-classical hybrid convolutional networks for classification which is
applied to the full image data of the second and third data sets. In either of
the two methodologies, the goal is a comparison of the performance between the
fully-classical approach with the one involving quantum circuits as part of the
total classification pipeline.

\subsection{Methodology 1: Randomized neural networks}
\label{subsec:method:rnd}
The first methodology, which acts on the dimensionality-reduced data sets, may
be viewed as a two-stage protocol in order to classify the various data sets
introduced in Sec.~\ref{sec:data}. However, the first stage must be viewed as
a mere necessity, as it is required to render the image data of the second and
third data set approachable for quantum neural networks --- an unpleasant but
unfortunately currently inevitable step that enables the comparison of classical
and quantum neural networks in the first place. It prepares the data such that
the actual classification, which formally happens in the second stage, only
receives data of moderate dimensionality which both the classical and quantum
methods can handle. We are aware that this first data reduction step, if viewed
independently from the second step, renders the classification task that is
being solved in the second step much simpler, since a lot of the
``heavy-lifting'' is already solved in the first step. This is especially
evident when viewing the reduced data for the handwritten digits from MNIST,
independently of whether a CAE or a PCA is used to reduce the original images,
cf.\ Fig.~\ref{fig:overview}~(a) and~(b), respectively. In that case, even the
dimensionality reduction to $d = 2$ features shows a very clear spatial
separation of $0$'s and $1$'s, which already suggests a rather simple
classification task for the second stage. In contrast, for the industrial images
from TRUMPF, a reduction to $d = 2$ features seems too drastic in order to allow
a proper reconstruction of the original image, cf.\ Fig.~\ref{fig:overview}~(q)
for a reduction via CAE although the results for a reduction via PCA are
qualitatively similar. This is moreover evidenced by the reduced 2D data shown
in Fig.~\ref{fig:overview}~(d) and~(e), where no clear spatial separation can be
observed. In that case, a larger reduced dimensionality $d$ is required, e.g.\
$d = 4$ or $d = 8$, to reproduce basic characteristics of the original image.
Unfortunately, their corresponding data cannot be visualized trivially.

Having motivated and discussed the necessity of the dimensionality reduction of
the first step, we now simply take its output data as input data for the second
stage. From a mathematical point of view, the second stage thus gets a set
$\mathcal{D} = \{(\myvec{x}_{i}, y_{i}^{\rmtrue})\}_{i = 1, \dots,
N_{\rmtrain}}$ of vectors $\myvec{x}_{i} \in \mathbb{R}^{d}$ and labels
$y_{i}^{\rmtrue} \in \{0, 1\}$, with $N_{\rmtrain}$ the number of training data
points, that follows an unknown classification function $f: \mathbb{R}^{d}
\rightarrow \{0, 1\}$ which needs to be learned --- or approximated --- by
a machine-learning model given only $\mathcal{D}$. To do that, we consider both
classical and quantum neural networks for classifying the data. In order to
allow for an as fair as possible comparison between classical and quantum
methods, we utilize a randomization scheme and generate classical and quantum
neural networks with randomized hyperparameters. Their details are specified in
Secs.~\ref{subsubsec:method:ClassNNs} and~\ref{subsubsec:method:QNNs},
respectively. Through this, we ensure that we explore a larger --- albeit not
exhaustive --- fraction of the individual hyperparameter spaces. As
a consequence, this allows us to not only compare the best performing classical
and quantum models generated this way, but also how average or badly performing
models, which typically result from a poor choice of hyperparameters, compare
between classical and quantum approaches.

In detail, we generate $50$ random classical and quantum neural networks for
each data set introduced in Sec.~\ref{sec:data}. Since a model's performance,
judged on the classification accuracy achieved on the validation data set, may
depend strongly on the model's initial parameters which are used during
training, we train each of the $50$ classical and quantum models with $10$ sets
of random initial parameters. Each of the $10$ sets is trained for up to $100$
epochs, which suffices for convergence in almost all cases. As usual, we define
one epoch as one complete pass through the training data set. This averaging
heavily diminishes the effect of having exceptionally good or bad luck when
choosing initial training parameters and shifts the focus to a model's average
performance when --- as in most cases --- no a-priori knowledge about good
initial trainable parameters is available. The performance of any of the $50$
random classical or quantum models is thus given by the average of the $10$
individual final classification accuracies achieved on the validation data set
after training on the training data set.

\subsubsection{Classical neural networks}
\label{subsubsec:method:ClassNNs}

In order to generate random classical neural network architectures that allow
processing of data vectors $\myvec{x}_{i} \in \mathbb{R}^{d}$, there are various
hyperparameters that can be randomly determined. Among the most prevalent ones
are the number of dense, i.e., fully-connected, layers and the number of neurons
per layer as well as the employed activation functions. We decide to randomize
the first two by randomly choosing $n_{\rmlayer, \rmc} \in \{1, 2, 3, 4\}$ dense
layers for each neural network, and independently choosing $n_{l, \rmc} \in \{2,
3, 4\}$ neurons for each dense layer at random, with $l = 1, \dots, n_{\rmlayer,
\rmc}$. Note that we actively decide to keep both numbers small in order to keep
the models simple and the number of trainable parameters of the ensuing
classical neural networks comparable to those of the quantum neural networks,
whose randomized architectures are introduced in
Sec.~\ref{subsubsec:method:QNNs}. While the number of trainable parameters is
only one factor among many defining the (expressive) power of machine-learning
models, with other factors being, for instance, their non-linearities, it is
nevertheless a vital factor and its equivalence a prerequisite for comparability
between different models --- i.e., between classical and quantum architectures
in our case. We thereby diminish the possibility that potentially better results
for either classical or quantum neural networks are due to the fact that one has
more trainable parameters available than the other. For the activation
functions, we restrict our study to architectures where we use
ReLU~\cite{Agarap2018}, defined as $\mathrm{ReLU}(x) = \max(0,x)$, between every
layer except for the final layer, which consists only of a single neuron for
classification that is connected to its previous layer via the sigmoid
activation function, defined as $\mathrm{\sigma}(x) = (1 + e^{-x})^{-1}$.

While there are many other choices and hyperparameters to vary, such as the
optimizer, the loss function or the learning rate, we decide to keep them fixed
for all classical models in order to allow better comparability with quantum
models~\footnote{This is primarily because the employed software for the quantum
neural networks~\cite{Kreplin2023} provides less options than that employed for
the classical neural networks~\cite{tensorflow2015-whitepaper}}. We choose the
Adam optimizer~\cite{Kingma2014}, the mean squared error
\begin{align}
  \mathcal{L}_{\mathrm{mse}}
  =
  \frac{1}{N_{\rmtrain}}
  \sum_{i=1}^{N_{\rmtrain}} \left(y_{i}^{\rmtrue} - y_{i}^{\rmpred}\right)^{2},
\end{align}
as the loss functions, with $y_{i}^{\rmtrue}$ and $y_{i}^{\rmpred}$ the true and
predicted label of the $i$th data point, respectively, and a learning rate
$R_{\rmc}^{(1)} = 0.005$.

\subsubsection{Quantum neural networks}
\label{subsubsec:method:QNNs}

In the following, we introduce how we generate random quantum neural network
architectures, which, in practice, are given by parametrized quantum circuits
(PQCs). To this end, note that their hyperparameters are --- at least partially
--- very different from those of classical neural networks.

In view of quantum machine learning via PQCs, the solution to classification
tasks, such as those described in Sec.~\ref{sec:data}, can typically be divided
into several steps, which we outline in the following before specifying our own
adapted approach afterwards.
\begin{enumerate}
  \item[(i)] First, the initial state $\ket{\Psi_{\rminit}} = \ket{0}^{\otimes
    n_{\rmq}}$, with $n_{\rmq}$ the number of qubits, is prepared.
  \item[(ii)] Afterwards, the classical data from the vector $\myvec{x}_{i} \in
    \mathbb{R}^{d}$ is encoded into the quantum state via a quantum feature map
    $\op{S}_{\rmenc}(\myvec{x}_{i})$, i.e., $\ket{\Psi_{\rmenc}(\myvec{x}_{i})}
    = \op{S}_{\rmenc}(\myvec{x}_{i}) \ket{\Psi_{\rminit}}$.
  \item[(iii)] Then, a parametrized unitary $\op{U}(\myvec{\theta})$, composed
    out of $M_{\rmparam}$ parametrized quantum gates, with $\myvec{\theta} \in
    \mathbb{R}^{M_{\rmparam}}$ the vector of gate parameters, is applied to
    yield $\ket{\Psi_{\rmout}(\myvec{x}_{i}, \myvec{\theta})}
    = \op{U}(\myvec{\theta}) \ket{\Psi_{\rmenc}(\myvec{x}_{i})}$.
  \item[(iv)] Finally, a measurement with a Hermitian operator $\op{O}$ is
    performed to yield the expectation value $y_{i}^{\rmpred}(\myvec{x}_{i},
    \myvec{\theta}) = \braket{\Psi_{\rmout}(\myvec{x}_{i}, \myvec{\theta})
    | \op{O} | \Psi_{\rmout}(\myvec{x}_{i}, \myvec{\theta})} \in \mathbb{R}$.
\end{enumerate}
The training aspect of this quantum-machine-learning ansatz is then to
iteratively optimize the parameter vector $\myvec{\theta}$ on a classical
computer such that $y_{i}^{\rmpred}$ coincides with the true label
$y_{i}^{\rmtrue}$ assigned by the unknown classification function $f:
\mathbb{R}^{d} \rightarrow \{0, 1\}$ for all data points in $\mathcal{D}
= \{(\myvec{x}_{i}, y_{i}^{\rmtrue})\}_{i = 1, \dots, N_{\rmtrain}}$. However,
if trained successfully, the model ideally generalizes to data points
$(\myvec{x}', y') \notin \mathcal{D}$ beyond its training data set, but
belonging to a source with the same distribution.

In order to adapt the above scheme to our approach of generating random quantum
neural network models, there are various hyperparameters that can be varied.
First, we allow for a random number $n_{\rmq}$ of qubits. In detail, to generate
a random quantum neural network able to process $d$-dimensional input vectors,
we choose $n_{\rmq} \in \{2, 4, 8, \dots, d\}$ randomly, exploiting that $d$ is
always a power of two. Second, and intimately connected to the choice of
$n_{\rmq}$, is the choice of the quantum feature map
$\op{S}_{\rmenc}(\myvec{x}_{i})$, which, inspired by Refs.~\cite{Lloyd2020,
PRA.106.042431, PRA.109.042421, PhysRevResearch.6.043326}, may become trainable as well in our case, i.e.,
$\op{S}_{\rmenc}(\myvec{x}_{i}, \myvec{\phi})$ with $\myvec{\phi} \in
\mathbb{R}^{M_{\rmenc}}$ a vector of $M_{\rmenc}$ trainable parameters. For
simplicity, we first discuss the case $n_{\rmq} = d$, since the case $n_{\rmq}
< d$ is more contrived. For $n_{\rmq} = d$, the quantum feature map is chosen
such that
\begin{align} \label{eq:qfm}
  \op{S}_{\rmenc}(\myvec{x}_{i}, \myvec{\phi})
  =
  \bigotimes_{n = 1}^{n_{\rmq}}
  \op{R}_{\alpha, n} \left(
    f_{\rmenc}(x_{i, n}, \phi_{n})
  \right)
\end{align}
where $\op{R}_{\alpha, n}(\beta) = \exp\{- \im \alpha \beta/2\}$ is a rotational
single-qubit gate, applied to the $n$th qubit, parametrized by an angle $\beta$
and specified by the Pauli operator $\alpha \in \{\op{X}, \op{Y}, \op{Z}\}$. The
latter is chosen randomly but identical for all $n_{\rmq}$ qubits. This encoding
scheme, i.e., to encode one feature $x_{i, n}$ per qubit, has been identified
previously as a viable embedding scheme~\cite{Grant2018, PRA.101.052309}. The
encoding function $\beta = f_{\rmenc}(x_{i, n}, \phi_{n})$ is also chosen
randomly from a set of predefined functions $f(x, \phi)$ encompassing the
functions
\begin{align}
  \left\{%
    x,
    \arccos(x),
    x + \phi,
    x \phi,
    \arccos(x) \phi
  \right\}.
\end{align}
Note that some functions are accompanied by an additional trainable parameter
$\phi$ while some just contain the feature value $x$. The inclusion of
$\arccos(x)$ as a non-linearity is inspired by Ref.~\cite{PRA.98.032309}. In
total, this randomized quantum feature map $\op{S}_{\rmenc}(\myvec{x}_{i},
\myvec{\phi})$ produces the encoded state $\ket{\Psi_{\rmenc}(\myvec{x}_{i},
\myvec{\phi})}$ to which we then apply a parametrized unitary
$\op{U}(\myvec{\theta})$ consisting of randomly chosen gates as follows. First,
we apply another randomized but identical single-qubit rotational gate
$\op{R}_{\alpha, n}(\beta)$ to each qubit $n = 1, \dots, n_{\rmq}$, where, this
time, $\beta = \theta$ is plainly a trainable parameter from $\myvec{\theta} \in
\mathbb{R}^{M_{\rmparam}}$. This is followed by a randomized choice of having no
entangling gates or any of the three controlled $\op{cR}_{\alpha}(\beta)$ gates,
$\alpha \in \{\op{X}, \op{Y}, \op{Z}\}$, defined via
\begin{align}
  \op{cR}_{\alpha}(\beta)
  =
  \ket{0} \bra{0}^{\rmctrl} \otimes \uop^{\rmtrgt}
  +
  \ket{1} \bra{1}^{\rmctrl} \otimes \op{R}_{\alpha}^{\rmtrgt}(\beta)
\end{align}
with superscripts ``$\rmctrl$'' and ``$\rmtrgt$'' denoting control and target
qubit, respectively. If the randomization chooses to have entangling gates, this
implies that several of them are being placed, with each of them having their own
trainable parameter. We choose randomly between two possible structures. On the
one hand, we may choose to have a linear structure in the sense that there is an
entangling gate between qubits $1$ and $2$, between $2$ and $3$ and so on until
qubits $n_{q}-1$ and $n_{\rmq}$. On the other hand, we may choose an all-to-all
connected structure, i.e., an entangling gate between any pair of
qubits. Note that the difference only matters for $n_{\rmq} > 2$. The
single-qubit rotational gates together with the randomized trainable entangling
gates make up the parametrized quantum circuit $\op{U}(\myvec{\theta})$.

As another choice of randomization and inspired by Refs.~\cite{PerezSalinas2020,
PRA.103.032430}, which motivate the repetition of the data encoding and the
parametrized circuit to enhance a model's expressivity, we randomly choose to
execute $n_{\rmlayer, \rmq} \in \{1, 2, 3, 4\}$ layers of the quantum feature
map and the parametrized unitary. Our randomized PQC (for $n_{\rmq} = d$) thus
reads
\begin{align} \label{eq:pqc:full}
  \op{U}_{\rmpqc}(\myvec{x}_{i}, \myvec{\phi}, \myvec{\theta})
  =
  \op{U}_{\rmhad}
  \prod_{l = 1}^{n_{\rmlayer, \rmq}} \Big[
    \op{U}(\myvec{\theta}_{l})
    \op{S}_{\rmenc}(\myvec{x}_{i}, \myvec{\phi}_{l})
  \Big],
\end{align}
where $\myvec{\phi}_{l}$ and $\myvec{\theta}_{l}$ are trainable parameters that
are independent for each layer $l$ despite $\op{S}_{\rmenc}(\myvec{x}_{i},
\myvec{\phi}_{l})$ and $\op{U}(\myvec{\theta}_{l})$ having the same overall
structure in each layer. As a final choice of randomizing the PQC, the
additional unitary $\op{U}_{\rmhad}$ randomly applies either a Hadamard gate or
the identity to all qubits. This is a means to rotate the initial state
$\ket{\Psi_{\rminit}} = \ket{0}^{\otimes n_{\rmq}}$ into a more accessible basis
prior to executing the remainder of the PQC.

Having so far discussed the case $n_{\rmq} = d$, we now discuss $n_{\rmq} < d$.
In those cases, the quantum feature map, as introduced in Eq.~\eqref{eq:qfm},
cannot be directly used since there are fewer qubits than there are features
that need encoding. However, we can choose the same structure as in the final
PQC of Eq.~\eqref{eq:pqc:full} and adapt it to have multiple quantum feature
maps that each encode a fraction of the total data vector $\myvec{x}_{i} \in
\mathbb{R}^{d}$. In detail, we adapt Eq.~\eqref{eq:pqc:full} to become
\begin{widetext}
\begin{align} \label{eq:pqc:red}
  \op{U}_{\rmpqc}(\myvec{x}_{i}, \myvec{\phi}, \myvec{\theta})
  =
  \op{U}_{\rmhad}
  \prod_{l = 1}^{n_{\rmlayer, \rmq}} \Bigg[
    \prod_{r = 1}^{n_{\rmrep}} \Big[
      \op{U}_{r}(\myvec{\theta}_{l, r})
      \op{S}_{\rmenc, r}(\myvec{x}_{i, r}, \myvec{\phi}_{l, r})
    \Big]
  \Bigg],
\end{align}
\end{widetext}
where $\myvec{x}_{i,r} \in \mathbb{R}^{n_{\rmq}}$ are fractions of
$\myvec{x}_{i} = (\myvec{x}_{i,1}, \dots, \myvec{x}_{i,n_{\rmrep}}) \in
\mathbb{R}^{d}$ with $n_{\rmrep} = d/n_{\rmq}$. Note that we allow the quantum
feature maps $\op{S}_{\rmenc, r}(\myvec{x}_{i, r}, \myvec{\phi}_{l, r})$ and the
parametrized unitary $\op{U}_{r}(\myvec{\theta}_{l,r})$ to be chosen differently
in every repetition $r = 1, \dots, n_{\rmrep}$, following the same randomization
procedure as introduced before. However, their randomly chosen architectures are
again identical for each layer $l = 1, \dots, n_{\rmlayer, \rmq}$. In contrast,
their trainable parameters $\myvec{\phi}_{l, r}$ and $\myvec{\theta}_{l, r}$ are
different in each repetition and layer.

To finally map the output state
\begin{align} \label{eq:psi:out}
  \ket{\Psi_{\rmout}(\myvec{x}_{i}, \myvec{\phi}, \myvec{\theta})}
  =
  \op{U}_{\rmpqc}(\myvec{x}_{i}, \myvec{\phi}, \myvec{\theta})
  \ket{\Psi_{\rminit}}
\end{align}
to a scalar, we need to specify the measurement operator $\op{O}$. Within our
randomization approach, we make it trainable as well and allow it to be one of
the two following operators at random,
\begin{subequations}
\begin{align} \label{eq:Opauli}
  \op{O}_{\mathrm{pauli}}(\myvec{\omega})
  =
  \omega_{1} \uop
  +
  \sum_{n = 1}^{n_{\rmq}} \left(
    \omega_{n,\rmx} \op{X}_{n}
    +
    \omega_{n,\rmy} \op{Y}_{n}
    +
    \omega_{n,\rmz} \op{Z}_{n}
  \right)
\end{align}
or
\begin{align} \label{eq:Oising}
  \op{O}_{\mathrm{ising}}(\myvec{\omega})
  =
  \omega_{1} \uop
  &+
  \sum_{n = 1}^{n_{\rmq}} \left(
    \omega_{n,\rmx} \op{X}_{n}
    +
    \omega_{n,\rmz} \op{Z}_{n}
  \right)
  \notag \\
  &+
  \sum_{\substack{n, m = 1 \\ n < m}}^{n_{\rmq}}
    \omega_{n m} \op{Z}_{n} \op{Z}_{m},
\end{align}
\end{subequations}
where $\myvec{\omega} \in \mathbb{R}^{M_{\rmop}}$ captures the $M_{\rmop}$
trainable parameters of $\op{O}_{\mathrm{pauli}}(\myvec{\omega})$ or
$\op{O}_{\mathrm{ising}}(\myvec{\omega})$. Ultimately, this yields
the expectation value
\begin{align} \label{eq:exp}
  y_{i}^{\rmpred}(\myvec{x}_{i}, \myvec{\phi}, \myvec{\theta}, \myvec{\omega})
  =
  \Braket{%
    \Psi_{\rmout}(\myvec{x}_{i}, \myvec{\phi}, \myvec{\theta}) |
    \op{O}(\myvec{\omega}) |
    \Psi_{\rmout}(\myvec{x}_{i}, \myvec{\phi}, \myvec{\theta})
  }.
\end{align}
The training process for the PQC is then to optimize the circuit parameters
$\myvec{\phi}$ and $\myvec{\theta}$ as well as the operator parameter
$\myvec{\omega}$ in order to match $y_{i}^{\rmpred}(\myvec{x}_{i}, \myvec{\phi},
\myvec{\theta}, \myvec{\omega})$ and $y_{i}^{\rmtrue}$ as best as possible for
all $N_{\rmtrain}$ data points in $\mathcal{D}$. Although this process is
intended to be carried out in a quantum-classical hybrid approach, as described
earlier, note that we carry out all our simulations on classical computers,
involving extensive simulations of quantum circuits on a high-performance
computing infrastructure. Furthermore, note that the randomization procedure
described so far has a small but non-vanishing probability of yielding
non-trainable models, for instance, by exclusively adding diagonal gates like
$\op{R}_{\rmz}$ and $\op{cR}_{\rmz}$ to the PQC unitary $\op{U}_{\rmpqc}$, which
thus becomes diagonal as well and produces $\ket{\Psi_{\rmout}(\myvec{x}_{i},
\myvec{\phi}, \myvec{\theta})} = e^{\im \varphi} \ket{\Psi_{\rminit}}$ with
$\varphi$ some phase. The latter is the only variable that will change while
training such a PQC, which invalidates such PQCs for practical machine-learning
tasks. We do not neglect these models per se but exclude them in any analysis
where they would otherwise distort the statistics.

We employ again the Adam optimizer, use the squared error
\begin{align}
  \mathcal{L}_{\mathrm{se}}
  =
  \sum_{i=1}^{N_{\rmtrain}} \left(y_{i}^{\rmtrue} - y_{i}^{\rmpred}\right)^{2},
\end{align}
as the loss function and the learning rate $R_{\rmq}^{(1)} = 0.05$~\footnote{Note that we use a larger learning rate for the quantum neural networks, compared to their classical counterparts, in order to achieve a similar convergence behavior given the different loss functions}.

\subsection{Methodology 2: Convolutional neural networks}
\label{subsec:method:conv}
As already mentioned when introducing the first methodology in the previous
Sec.~\ref{subsec:method:rnd}, the employed dimensionality reduction is a mere
necessity --- but one that heavily alters the classification problem being
solved in its second stage. For our second methodology, we therefore avoid such
dimensionality reduction methods and directly classify the full image data.
However, while this is feasible for fully-classical models, it is still not
feasible for fully-quantum models. This is why we now resort to a comparison
between fully-classical and quantum-classical hybrid models, although in a way
that dimensionality reduction is not needed. The hybrid models considered for
this methodology consist of hybridizations of convolutional neural networks with
quantum neural networks used as analogs of their convolution filters, which in
classical convolutional layers are simple matrices. This hybrid model structure
was first explored in Ref.~\cite{Henderson2019}. One should not confuse these
models with the ones referred to as quantum convolutional neural networks
(QCNNs)~\cite{Cong2019}, which are essentially fully-quantum neural networks
that progressively trace out qubits in the middle of the circuit using, for
instance, mid-circuit measurements. The latter are responsible for preventing
barren plateaus in these models~\cite{PRX.11.041011} and also add non-linearity
to the quantum operations. However, they also make QCNNs classically simulable~\cite{Bermejo2024}. Since the hybrid models used are much more
computationally expensive to train due to many applications of each quantum
circuit, we opt to preselect a smaller set of combinations of hyperparameters
instead of relying on the randomization approach used before. However, we still
train each model with $10$ different sets of random initial parameters for $100$
epochs to ensure that the results are not biased by the choice of initial parameters.

We compare these hybrid models to fully-classical convolutional neural networks, which are a standard tool for image classification tasks. To ensure a fair comparison, we also preselected a set of combinations of hyperparameters for these models.
\xchange{Since the convolution operation is primarily used for image data, we do not use the linear-split data set, cf. Sec.~\ref{subsec:data:linsplit}, to benchmark these models.}
\change{
We opted to not include the linear-split data set, cf. Sec.~\ref{subsec:data:linsplit}, in our convolutional neural network experiments. This decision stems from a fundamental mismatch between the inductive biases of convolutional neural networks and the mathematical structure of the hypercube classification problem. Convolutional operations are designed to exploit spatial locality, translation invariance and hierarchical feature composition. These properties are naturally present in image data, as we further describe in Sec.~\ref{subsec:method:limits}. However, the linear split of a hypercube represents a purely geometric problem where features lack meaningful spatial relations. In fact, each dimension of the hypercube corresponds to an independent coordinate rather than a spatially correlated pixel, making convolution operations conceptually inappropriate for this task. Thus, including this data set could potentially lead to misleading conclusions about the relative performance of classical and quantum approaches.
}

The details of the used classical and hybrid
convolutional neural networks are specified in
Sec.~\ref{subsubsec:method:ClassCNNs} and Sec.~\ref{subsubsec:method:QCCNNs},
respectively.

\subsubsection{Classical convolutional neural networks}
\label{subsubsec:method:ClassCNNs}

Classical convolutional neural networks, in the following referred to as CNNs,
which we use in this second methodology with the full, non-reduced image data
sets, allow for the processing of data matrices $\myvec{X}_{i} \in
\mathbb{R}^{n_h\times n_w}$, where $n_h$ and $n_w$ are the number of pixels in
the height and width of the corresponding image, respectively.

The key distinction between CNNs and classical neural networks is the exchange
of simple matrix multiplications in dense layers --- which are linear operations
--- for discrete convolutions in convolutional layers --- which are non-linear.
A discrete convolution operation is defined as
\begin{align} \label{eq:conv}
  (\myvec{X} \ast \myvec{K})_{i,j}
  =
  \sum_{m = 0}^{k_h-1} \sum_{n = 0}^{k_w-1}
  \myvec{X}_{i+m, j+n} \myvec{K}_{m,n},
\end{align}
where $\ast$ denotes the convolution operation, $\myvec{X}$ is the input matrix,
$\myvec{K}$ is the filter matrix, the indices $i$ and $j$ denote the entry $(i,
j)$ of the output matrix, and $k_h$ and $k_w$ are the dimensions of the filter
matrix. The entries of the filter matrix are the trainable parameters of the
convolutional layer, which can have multiple filters. The output of the
convolution operation represented in Eq.~\eqref{eq:conv} is a matrix with
dimensions $n_h\times n_w$, the same as the input matrix. However, in CNNs, the
filter matrix can be applied to only some $k_h \times k_w$ submatrices of the
input matrix, with a stride $s$ determining the step size between each
application of the filter along both image axes. Then, the output of the
convolution operation is a matrix with dimensions $d_{f, c}^{\,h} \times d_{f,
c}^{\,w}$, where $d_{f, c}^{\,h} = (\hat{n}_h - k_h)/s + 1$ and $d_{f, c}^{\,w}
= (\hat{n}_w - k_w)/s + 1$. Here, $\hat{n}_h$ and $\hat{n}_w$ are the dimensions
of the input matrix after padding it, for instance, with zeros to ensure that
the filter can be applied to the edges of the original image.

Some of the typical hyperparameters which can be varied in CNNs are the number
of convolutional layers, the number of filters per convolutional layer, the size
of these filters, the stride of each convolution, and the use of additive biases
in each convolution operation. The latter are trainable parameters which are
added to the output of a convolution operation.

To mimic the chosen structure for the hybrid CNNs, described in the following
Sec.~\ref{subsubsec:method:QCCNNs}, we opted to use CNNs consisting of one usual
convolutional layer followed by a varying number $n_\mathrm{dconv}$ of
depth-wise convolutional layers. These depth-wise convolutional layers differ
from usual convolutional layers for input images with multiple channels, i.e.,
with a third dimension $d_3 > 1$. The depth-wise convolutional layers convolve
a different filter matrix with each output channel of the previous layer,
instead of applying each filter to the entire input image, which effectively
makes the filters 3-dimensional, with the third dimension equal to $d_3$. Thus,
applying usual convolutions after the first convolutional layer would require
$3$-dimensional filters if the previous layer used more than one filter, which
would increase the number of trainable parameters significantly. This choice
was again made to mimic the structure of the hybrid CNNs. The outputs of the
convolutional layers are flattened into a vector and processed by a dense layer
with two neurons and a softmax activation function, defined as
$\mathrm{softmax}(\myvec{x})_i = e^{x_i} / \sum_{j=1}^{n} e^{x_j}$, where $i$ is
the index of the class for which we calculate the softmax value and $n=2$ is the
number of classes in our problem. The softmax function is used to transform the
outputs into a probability distribution over the predicted output classes.

To mimic the hyperparameter sets chosen for the hybrid CNNs, we opted for using
filters with sizes $2 \times 2$ or $3 \times 3$ for all convolutional layers.
For the first convolutional layer, we chose strides equal to the size of the
filters, such that each filter is applied once to each image pixel. Depending on
the choice of filter size, we use either four or nine filters in the first
convolutional layer, respectively. For the depth-wise convolutional layers, we
convolve each output channel of the previous layer with a single filter and
a stride of step size $s = 1$, to guarantee that the dimension of the input to
the dense layer remains the same whether we use depth-wise convolutional layers
or not. This way, we can focus our analysis on the number of trainable
parameters in the convolutional layers, since the dense layers will have the
same number of parameters, regardless of the number of convolutional layers we
use. In order to have between one and three convolutional layers, we tested the
number of depth-wise convolutional layers $n_\mathrm{dconv} \in \{0, 1, 2\}$. We
also decided to test the introduction of additive biases in all convolutional
layers at once.

Since we fix the model structures in the second methodology, it is simple to
calculate the number of trainable parameters in these networks formally.
Therefore, the considered CNNs have a total number of trainable parameters
$M_{\mathrm{cnn}}$ given by
\begin{subequations}
\begin{align}
  M_{\mathrm{cnn}}
  &=
  M_{\mathrm{conv}} + M_{\mathrm{dense}},
  \\
  M_{\mathrm{conv}}
  &=
  \left[(k_h k_w)^2 + \delta_b k_h k_w\right] l,
  \\
  M_{\mathrm{dense}}
  &=
  \frac{2 \hat{n}_h \hat{n}_w k_h k_w}{s^2} + 2,
\end{align}
\end{subequations}
where $M_{\mathrm{conv}}$ is the number of parameters in the convolutional
layers, $M_{\mathrm{dense}}$ is the number of parameters in the final dense
layer, and $\delta_b \in \{0,1\}$ indicates if the convolutions have additive
biases or not.

We again use the Adam optimizer with a learning rate
$R_{\rmc}^{(2)} = 0.01$ and the cross entropy
\begin{align}
  \mathcal{L}_{\mathrm{ce}}
  =
  -
  \sum_{i=1}^{N_{\rmtrain}}
  \sum_{j=1}^{n_{\mathrm{class}}}
  y_j^{\rmtrue} \log \left( y_{i,j}^{\rmpred} \right),
\end{align}
as the loss function, where $n_{\mathrm{class}}=2$ is the number of classes in
our problem, $y_j^{\rmtrue}$ represents the class $j$, in our case
$y_1^{\rmtrue}=0$ and $y_2^{\rmtrue}=1$, and $y_{i,j}^{\rmpred}$ is the
predicted probability of the $i$th input image belonging to class $j$. This loss
function measures the difference between the predicted probability distribution
and the true distribution of the labels in the training data set.

\subsubsection{Quantum-classical hybrid convolutional neural networks}
\label{subsubsec:method:QCCNNs}

As mentioned before, the quantum-classical hybrid convolutional neural networks,
in the following referred to as QCCNNs --- for ``quantum-classical CNNs'' ---
use PQCs instead of classical convolutional layers to process the input data.
While the structure of these QCCNNs is inspired by Ref.~\cite{Henderson2019},
the quantum circuits we use are fully trainable instead of fixed.

\begin{figure}
  \centering
  \includegraphics[width=\columnwidth]{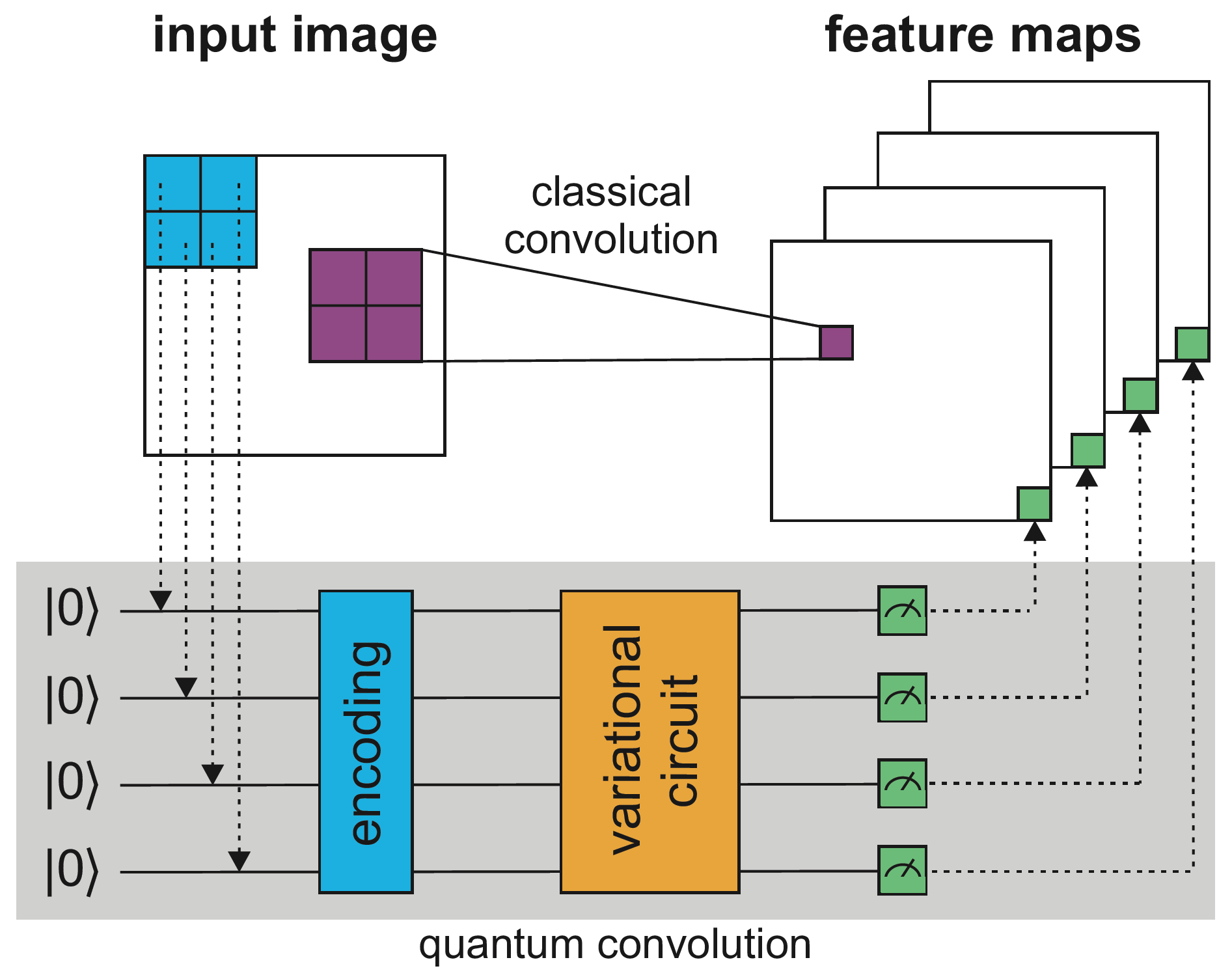}
  \caption{%
  Example scheme of a quantum convolution operation using a filter with
  dimensions $2\times2$ and a PQC with four qubits. In our QCCNN models, the
  input matrices are the non-reduced images, either from the handwritten-digit
  or the industrial-image data sets, cf.\ Secs.~\ref{subsec:data:mnist}
  and~\ref{subsec:data:trumpf}, and the output feature maps are flattened into
  a vector and fed to a classical dense layer. A minimal representation of an
  analogue classical convolution with the same filter size is also shown. Image
  adapted from Ref.~\cite{Matic2022}.
  }
  \label{fig:quantum_convolution}
\end{figure}

The structure of the studied QCCNNs is the same as that of the CNNs, cf.\
Sec.~\ref{subsubsec:method:ClassCNNs}, with the exception that all convolutional
layers are replaced by a single quantum convolution. The latter is defined as
a direct analogue of the classical discrete convolution, whereby the input
matrix is ``quantum convolved'' with a PQC by encoding the pixels of each $k_h
\times k_w$ submatrix of the input matrix --- defined by the ``quantum-filter''
size and stride $s$ of the (quantum) convolution as in the classical case ---
into the PQC and measuring the expectation value of an observable on the chosen
qubits of the output state. The output of the quantum convolution is
a $3$-tensor of expectation values of observables, whose first two dimensions
are dependent on the size of the input matrix and the third dimension is
dependent on the number of measured qubits in the PQC. This output has
dimensions $d_{f,q}^{\,h} \times d_{f,q}^{\,w} \times n_{\rmq, \mathrm{meas}}$,
where $d_{f,q}^{\,h} = d_{f,c}^{\,h}$, $d_{f,q}^{\,w} = d_{f,c}^{\,w}$ (cf.\
Sec.~\ref{subsubsec:method:ClassCNNs}), and $n_{\rmq, \mathrm{meas}}$ is the
number of measured qubits. Figure~\ref{fig:quantum_convolution} illustrates an
example of the quantum convolution operation described above, with four qubits,
as well as how it compares to a classical convolution. Note that, in this
example, we would need four classical convolutional filters to obtain an output
of the same dimensions as the one with the quantum convolution. The output of
the quantum convolution is then flattened into a vector, which is then
classified by classical means, i.e., by a dense layer of two neurons and
a softmax activation function, similar to the fully-classical case.

Considering a stride of $s=1$ for simplicity, the formula for the $n_{\rmq,
\mathrm{meas}}$-vector at position $(i, j)$ of the output $3$-tensor of
a quantum convolution with a quantum filter with dimensions $k_h \times k_w$, as a
function of its PQC, is
\begin{align}
  (\myvec{X} \ast U_{\rmpqc, \myvec{\phi}, \myvec{\theta}})_{i,j}
  =
  U_{pqc} \left(
    \myvec{X}_{i,i+k_h-1, j,j+k_w-1}, \myvec{\phi}, \myvec{\theta}
  \right),
\end{align}
where $\myvec{\phi}$ represents a tensor of all trainable parameters used in the
quantum feature map of the PQC, $\myvec{\theta}$ represents a tensor of all gate
parameters used in the following trainable unitaries, and $\myvec{X}_{i,i+k_h-1,
j,j+k_w-1}$ is the submatrix of $\myvec{X}$ which can be defined by its upper
left corner at position $(i, j)$ and its lower right corner at position
$(i+k_h-1, j+k_w-1)$, thus having the dimensions of the quantum filter.

For the QCCNNs, we also use quantum-filter sizes of $2 \times 2$ and $3 \times
3$ with strides equal to the size of the filters, i.e., $s = 2$ and $s = 3$,
respectively. To simplify the PQC structures, we encode only one feature per
qubit, such that the width $n_q$ of the PQC is defined by the quantum filter
size, $n_q = k_h k_w$, i.e., $n_{q} = 4$ for $2 \times 2$ filters and $n_{q}
= 9$ for $3 \times 3$ filters.

Since we fix a smaller set of combinations of hyperparameters, we need to ensure
we benchmark models which are still expressive enough. To this end, with the
insight that the choice of the quantum feature map
$\op{S}_{\rmenc}(\myvec{x}_{i}, \myvec{\phi})$ directly lower-bounds the
smallest training error achievable~\cite{PRA.110.022411}, we opt for a very
expressive trainable feature map of the form
\begin{widetext}
  \begin{align}
    \op{S}_{\rmenc}(\myvec{x}_{i}, \myvec{\phi})
    =
    \bigotimes_{n = 1}^{n_{\rmq}}
    \op{R}_n \left(
      f_{\rmenc}(x_{i, n}, \phi_{1,n}),
      f_{\rmenc}(x_{i, n}, \phi_{2,n}),
      f_{\rmenc}(x_{i, n}, \phi_{3,n})
    \right),
  \end{align}
\end{widetext}
where $\op{R}_{n}(\alpha, \beta, \omega) = \op{R}_{\op{Z},n}(\omega)
\op{R}_{\op{Y},n}(\beta) \op{R}_{\op{Z},n}(\alpha)$ (cf.\
Sec.~\ref{subsubsec:method:QNNs} for the definition of the individual gates) is
a general single-qubit rotation gate applied to the $n$th qubit and parametrized
by the three angles $\alpha, \beta, \omega$. The encoding function $f_{\rmenc}$
is fixed by the non-linear form $f_{\rmenc}(x, \phi) = \arccos(x) \phi$ for
all three angles. While we encode each pixel $x_{i, n}$ in all angles of this
general rotation, each angle has its own trainable parameter, $\phi_{j,n},
j \in \{1,2,3\}$. This quantum feature map receives as input a submatrix
flattened into a vector $\myvec{x}_i \in \mathbb{R}^{n_{q}}$, and a matrix of
trainable parameters $\myvec{\phi} \in \mathbb{R}^{3 \times n_q}$.

We decided to test QCCNNs with one of two entangling structures, namely circular
entanglement, which is similar to the linear structure described in
Sec.~\ref{subsubsec:method:QNNs} but with an additional entangling gate between
qubits $n_q$ and $1$; and all-to-all entanglement, described in the same
Section. We fix the entangling gates used in all QCCNNs with the form
$\op{cR}_{\op{X}}(\theta)$, with $\theta$ being a trainable parameter specific to
each gate. Additionally, we also test QCCNNs without entangling gates to
benchmark the usefulness of entanglement in these hybrid models for the
classical data sets tested.

As with the approach based on randomized quantum neural networks defined in
Sec.~\ref{subsubsec:method:QNNs}, we define a quantum layer as the quantum
feature map followed by trainable entangling gates. In practice, we test PQCs
with $n_{\rmlayer, \rmq} \in \{1, 2, 3\}$ quantum layers. This is a direct
analogy of the convolutional-layer structure used in the classical CNNs.
However, in contrast to the measurement operators employed by the randomized
quantum neural networks, cf.\ Eqs.~\eqref{eq:Opauli} and~\eqref{eq:Oising}, we
fix the measurement observable for each qubit as the Pauli $\op{Z}$ operator.

The PQCs we use in the QCCNNs take the form
\begin{align}
  \op{U}_{\rmpqc}(\myvec{x}_{i}, \myvec{\phi}, \myvec{\theta})
  =
  \prod_{l = 1}^{n_{\rmlayer, \rmq}} \Big[
    \op{U}(\myvec{\theta}_{l})
    \op{S}_{\rmenc}(\myvec{x}_{i}, \myvec{\phi}_{l})
  \Big],
\end{align}
where $\op{U}(\myvec{\theta}_{l})$ represents the unitary describing the
application of the parametrized entangling gates after the quantum feature map
for each layer $l$ of the PQC.

The QCCNNs described so far have a number of trainable parameters given by
\begin{subequations}
\begin{align}
  M_{\mathrm{qccnn}}
  &=
  \left(M_{\rmenc} + M_{\rment}\right) n_{\rmlayer, \rmq} + M_{\mathrm{dense}},
  \\
  M_{\rmenc}
  &= 3 k_h k_w,
  \\
  M_{\mathrm{dense}}
  &=
  \frac{2 \hat{n}_h \hat{n}_w k_h k_w}{s^2} + 2,
\end{align}
\end{subequations}
where $M_{\rmenc}$ is the number of parameters in one application of the quantum
feature map, $M_{\rment}$ is the number of parameters in the entangling gates of
each layer and $M_{\mathrm{dense}}$ is the number of parameters in the final
dense layer. For QCCNNs with circular entangling gates, $M_{\rment} = k_h k_w$,
while for those with all-to-all entangling gates, $M_{\rment} = [k_h k_w (k_h
k_w-1)]/2$. Obviously, for QCCNNs without entangling gates, $M_{\rment} = 0$.

We highlight that the CNNs and QCCNNs we consider have a very similar number of
parameters, only differing slightly in the number of parameters in their
convolutions. For the scenarios we consider in the following, specifically for
the handwritten-digit data set with $2\times 2$ ($3\times 3$) convolution
filters, the dense layer has $M_{\mathrm{dense}} = 1570\ (1802)$ parameters,
with the difference originating from image padding. However, due to the much
larger dimensions of the industrial images, this data set brings the number of
parameters of the dense layer to $M_{\mathrm{dense}} \approx 32000$. In
fairness, most of the parameters of both CNNs and QCCNNs belong to their final
dense layer. Thus, we find it reasonable to benchmark all models against
a single classical dense layer with two neurons and softmax activation function
as a baseline. Moreover, we note the number of parameters in the models of the
second methodology is much larger than in the randomized models of the first
methodology. The presumable increase in the models expressivity, compared to
that of the first methodology, is necessary for them to be able to extract
useful information from the original and non-reduced $784$ or $16000$ image
features. This is in contrast to the first methodology where models need to
extract information from at most $64$ features.

Given the similar structures between classical and hybrid CNNs, we decide to use
the same optimizer, learning rate $R_{\rmq}^{(2)} = R_{\rmc}^{(2)}$, and loss
function as with classical CNNs, cf.\ Sec.~\ref{subsubsec:method:ClassCNNs}.

As mentioned before, these hybrid models are computationally expensive due to
the many applications of the PQC in the quantum convolution. For instance, for
a $2\times 2$ quantum filter with stride $s=2$, each image of the
handwritten-digit data set requires $196$ applications of the $4$-qubit PQC per
iteration. For the industrial data set in the same scenario, this figure
increases to $4000$. Hence, since all the models were simulated on classical
hardware, this study was only possible through the use of high-performance
computing infrastructure paired with just-in-time (JIT) compilation of our code.
Even then, the longest simulation of a QCCNN still took $\sim 5$ days for each
set of random initial parameters. This was the case for the QCCNN with
a $3\times 3$ quantum filter, stride of $s=3$, and a $9$-qubit PQC with
all-to-all entanglement and $n_{\rmlayer, \rmq} = 3$ layers, applied on the
industrial-image data set with $N=500$ data points.

\change{%
\subsection{Limits of the methodologies}
\label{subsec:method:limits}
In the following, we want to briefly summarize our methodologies and point out possible limitations in order to provide a clear picture to the reader.
}

\change{%
For the first methodology, cf. Sec.~\ref{subsec:method:rnd}, we use randomized classical and quantum neural networks to classify dimensionality-reduced data sets, cf. Sec.~\ref{sec:data}. Since the dimensionality reduction is done via fully-classical methods, either by PCA or CAE, comparing the performances of classical and quantum neural networks in classifying the data becomes a comparison between fully-classical and quantum-classical hybrid methods. Moreover, to ensure a fairer comparison, we restrict the size of the classical neural networks such that they have a similar number of trainable parameters compared to their quantum counterparts. For the quantum neural networks, restricting them to modest sizes is necessary, on the one hand, to keep them in the feasible computational regime of current and near-term quantum hardware, and, on the other hand, to keep them classically simulable. At last, we note that we generate neural networks randomly instead of performing a structured hyperparameter search. This allows us to explore a larger fraction of the hyperparameter space.
}

\change{%
In the second methodology, cf. Sec.~\ref{subsec:method:conv}, we avoid dimensionality reductions via PCA or CAE for the image data, but apply classical or quantum convolutional layers on the full image data. Afterwards, its output is transferred to a classical dense layer responsible for the final classification. Hence, this methodology is a comparison between fully-classical and quantum-classical hybrid methods as well. Moreover, this methodology is numerically expensive to simulate, which is why we avoid randomization of the convolutional layer structures, but restrict ourselves to some preselected ones. In contrast to the first method, we do not use optimizable measurements or randomized types of entangling gates. In this case, only the structure of the entangling gates is randomized.
}

\change{%
The literature suggests that optimal performance often depends on the alignment between a model's inherent assumptions --- its inductive biases --- and the underlying structure of the data \cite{Gili2024}.
Each methodology has its associated inductive biases and is suited for different types of data or learning objectives.
In our first methodology, the randomized hyperparameter approach diversifies these biases.
Nevertheless, the models used are still relatively simple.
Both the classical and quantum neural networks incorporate minimal inductive biases beyond basic function approximation, hopefully allowing them to learn relatively unconstrained representations of the dimensionality-reduced data.
At the same time, this means that the models may not be able to fully exploit the underlying structure of each data set, especially the more complex industrial image data set.
}

\change{%
On the other hand, the CNNs in our second methodology have strong inductive biases toward spatial locality, translation invariance, and hierarchical feature composition, as described in Sec.~\ref{subsec:method:conv}.
These biases are fundamentally designed with images in mind.
They aim to exploit the assumption that neighboring pixels often contain related information and that similar useful patterns can appear at different spatial locations.
QCCNNs inherit these spatial biases from their convolutional-like procedure, while introducing different quantum-specific inductive biases through their PQCs.
Hence why, in this methodology, we focus on image data sets.
}

\change{%
Moreover, the dimensionality reduction methods used in the first methodology, PCA and CAE, favor different inductive biases themselves. Since this can further skew the benchmark results for general binary image classification tasks, we opted to use both methods in our analysis.
}

\change{%
Nevertheless, we must note that we did not consider the specific inductive biases of each generated model, especially in the first methodology. Instead, we treat quantum architectures as black boxes which we explore through randomization. Thus, our methodology distinguishes itself from other works that specifically design PQCs to exploit known structures in the data, such as linearly conserved quantities or data-ordering effects~\cite{Bowles2023, Gili2024}.
}

\change{%
For both methodologies, we assess model performance with a focus on their classification accuracies after $100$ epochs of training, even though an argument can be made for comparing other performance metrics, such as the convergence speed for similar accuracies or the robustness to different hyperparameter choices. 
The learning rates are also not changed or optimized. While we tested various learning rates initially, we did not observe a noticeable impact on the model's final classification accuracy after $100$ training epochs. Hence, we fixed the rates to those mentioned before as they reliably ensure convergence of the classification accuracy in almost all cases. At last, we note that we did not include noise in the quantum models even though this will most likely lead to lower performance of the quantum models.
}

\section{Results from random neural networks}
\label{sec:results:rnd}
In this section, we exclusively present and discuss the results obtained by
employing the methodology based on randomized classical and quantum neural
networks, introduced in Sec.~\ref{subsec:method:rnd}, and applied to all data
sets from Sec.~\ref{sec:data}.

\subsection{Comparison of classical and quantum neural network performances}
\label{subsec:results:rnd:CvsQ}

\begin{figure*}
  \centering
  \includegraphics{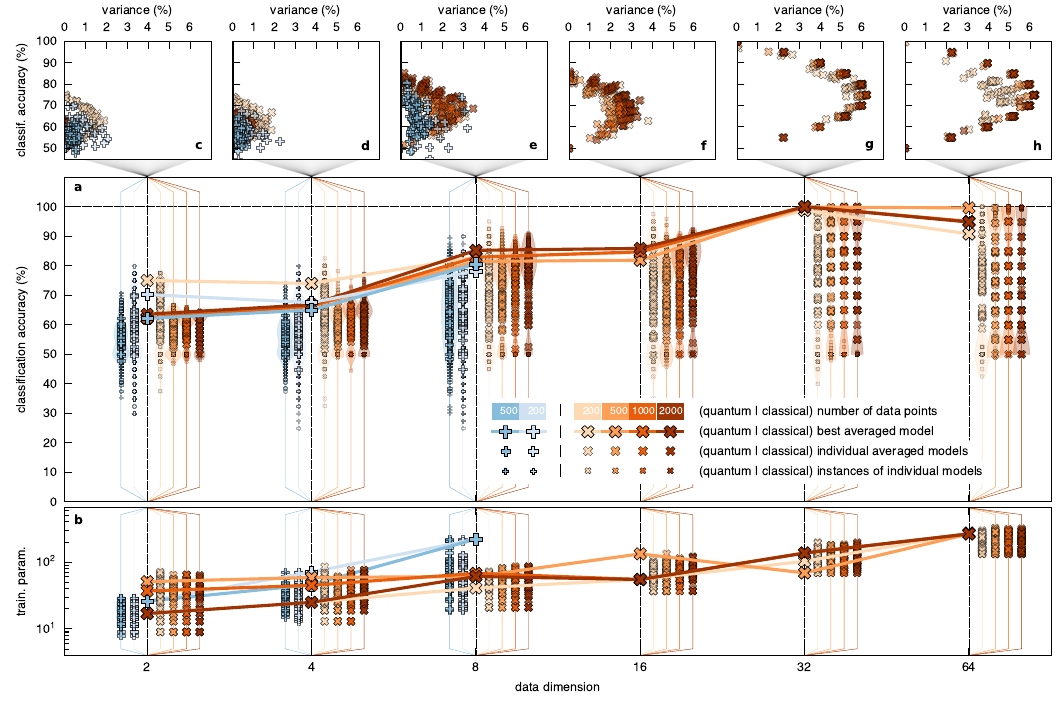}
  \caption{%
    Results for the third data set, i.e., the industrial image snippets from
    laser cutting machines, using a PCA for dimensionality reduction. Panel (a)
    shows the performance of $50$ random classical (orange tones) and quantum
    (blue tones) neural networks for different data dimensions, i.e., sizes of
    data reduction, and data set sizes, ranging from $200$ to $2000$ data points
    split into training and validation data with a ratio of $80\%$ to $20\%$.
    The medium sized markers show the $50$ average performances of each of these
    $50$ random models where the average is taken with respect to the outcome
    from training with $10$ sets of random initial parameters, cf.\
    Sec.~\ref{sec:method}. The outcome from each of these sets is depicted by
    the smallest markers. The largest markers, connected additionally by
    a line for better visibility, indicate the performance of the best of the
    $50$ random models. Panel (b) shows the corresponding number of trainable
    parameters for each model with those connected by a solid line reflecting
    the best performing models from panel (a). Panels (c)-(h) illustrate the
    different variances of the $50$ random models with respect to the $10$
    resulting accuracies obtained from the $10$ sets of random initial
    parameters.
  }
  \label{fig:results_dataset_3_pca}
\end{figure*}

\begin{figure}
  \centering
  \includegraphics{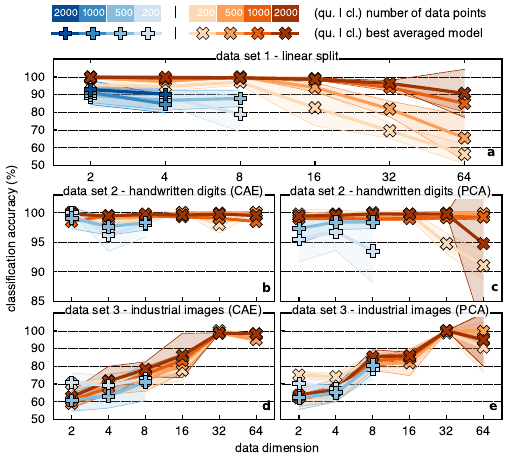}
  \caption{%
    Performance comparison between classical and quantum neural networks in
    terms of the classification accuracy of the best random models, i.e., the
    connected lines of Fig.~\ref{fig:results_dataset_3_pca}~(a) but for all data
    sets. The shaded background indicates the standard deviation generated by
    the $10$ individual random initial parameter sets that make up the shown
    averaged best random model performance.
  }
  \label{fig:results_joint}
\end{figure}

\change{%
\subsubsection{Detailed discussion of data set 3 --- industrial images --- under PCA reduction}
}

\change{\textit{Classical models} ---}
We start discussing our results by immediately comparing the performances of the
$50$ random classical and quantum neural networks in their ability to correctly
classify the various versions of the industrial image data set, cf.\
Sec.~\ref{subsec:data:trumpf}, since we believe them to be the most important
ones. We start with the various versions distinguished by their number of data
points $N \in \{200, 500, 1000, 2000\}$ and their reduced dimensionality $d \in
\{2, 4, 8, 16, 32, 64\}$ employing PCA as the reduction method. The results for
the first and second data sets, as well as the third data set when employing
a CAE as the dimensionality reduction method, are discussed afterwards.

Figure~\ref{fig:results_dataset_3_pca}~(a) shows the classification results for
each of the $50$ randomly generated classical neural networks in orange colors.
For the simplest classification task with $d = 2$, our randomization methodology
generates no models which perform excellently, i.e., yielding averaged
classification accuracies close to $100\%$. In fact, even the best models do not
surpass classification accuracies of around $\sim 75\%$ while the worst ones are
just slightly above $50\%$. There is, however, a notable difference depending on
the size of the training data set. Unless for the smallest one with $N = 200$, the
larger data sets yield at best classification accuracies at around $60\%$. Given
the rather drastic data reduction down to $d = 2$ features, the poor overall
classification accuracies are no surprise and the slightly better performance in
case of $N = 200$ may be attributed to an artificially less complex task determined by the small validation data set.
Figure~\ref{fig:overview}~(i) exemplarily shows the data points as well as the
learned function that classifies them best for $d = 2$ and $N = 500$~\footnote{\change{Note that the best performance of $\sim 75\%$, as mentioned earlier, refers to the averaged classification accuracies of the best model and $N = 200$. This should not be confused with the visual representation in Fig.~\ref{fig:overview}~(i), which corresponds to the best individual performance from among all $10$ sets of initial training results of the best averaged model and is moreover for $N = 500$}}. The
corresponding classical neural network architecture is illustrated directly
below Fig.~\ref{fig:overview}~(i). Upon moving towards data reductions up to $d
= 16$, we observe an increase in the best model's averaged classification
accuracies to roughly $80\%$, despite none of the individual runs with one set
of initial training parameters ever reaching $100\%$. For $d = 32$ and beyond,
our randomization approach starts yielding models giving rise to averaged
classification accuracies up to $100\%$, indicating that a reduction to $d = 32$
seems to capture enough information to warrant excellent classification.

\change{\textit{Quantum models} ---}
The various blue colors in Fig.~\ref{fig:results_dataset_3_pca}~(a) show the
classification results for the $50$ randomly generated quantum neural networks.
Due to their demanding computational cost for simulating and training them on
classical computers, we only generate networks that classify the data up to
dimension $d = 8$ and for data sets with $N \in \{200, 500\}$. We observe that
the maximally achieved classification accuracies are mostly similar to those
achieved by the $50$ randomly generated classical neural networks. At maximum,
we observe averaged classification accuracies of $\sim 70\%$ for $d = 2$ and of
$\sim 80\%$ for $d = 8$ for the best quantum models. The comparable performance
of classical and quantum models is despite the fact that many of the quantum
models perform on average worse than their classical counterparts. This is
indicated by the violin shapes in Fig.~\ref{fig:results_dataset_3_pca}~(a),
which indicate the bulk of the models' performances. Similar to
Fig.~\ref{fig:overview}~(i), which exemplarily shows the learned function of the
best performing classical neural network for $d = 2$ and $N = 500$,
Fig.~\ref{fig:overview}~(n) shows the same results for the best performing
quantum neural network. The latter is illustrated directly above
Fig.~\ref{fig:overview}~(n). By comparing Figs.~\ref{fig:overview}~(i) and~(n),
it is evident that despite the underlying data's complexity, the classical model
seems to approximate a rather simple function compared to the significantly more
complicated one chosen by the quantum model. In addition,
Fig.~\ref{fig:results_dataset_3_pca}~(b) shows the number of trainable
parameters for both classical and quantum models and confirms that both
architectures have roughly similar numbers~\footnote{An exact equivalence is not
possible due to the underlying randomization scheme in generating the models}.
Figure~\ref{fig:results_dataset_3_pca}~(b) also reveals that especially for the
quantum models, the ones with the largest number of trainable parameters tend to
perform best.

\change{\textit{Variances} ---}
In Fig.~\ref{fig:results_dataset_3_pca}~(c)-(h), we additionally analyze the
variances of the classification accuracies given the $10$ sets of random initial
training parameters whose after-training performance on the validation data set
defines the averaged classification accuracies shown in
Fig.~\ref{fig:results_dataset_3_pca}~(a). It can be seen that when comparing
results of classical and quantum models with similar averaged classification
accuracies, their corresponding variances differ slightly. In fact, the quantum
model variances seem to be generally smaller than those of the classical
models. This is especially evident in Fig.~\ref{fig:results_dataset_3_pca}~(e).
Thus, although the quantum neural networks perform similarly at maximum but worse
on average than their classical counterparts in terms of classification
accuracies, they yield those results more consistently and do depend less on the
initial starting parameters used for training. While a deeper statistical analysis is needed to confirm the generality of this property, this can be advantageous if good initial training parameters are not known in advance.

Moreover, we generally observe an increase in the accuracy variance of classical
models as we increase the number of data points in the data set. This can be due
to the additional data points increasing the separation hardness between the
classes and leading to more complex loss landscapes. While the same was not
observed for quantum models, this effect is clearer with higher data dimensions,
where we lack information about the performance of the latter due to the high
computational cost. Therefore, a deeper analysis would be needed for a fair
comparison but it might be an explanation for the surprisingly good performance
of the quantum model with $N = 200$ and $d = 2$ in
Fig.~\ref{fig:results_dataset_3_pca}~(a).

\change{%
\subsubsection{Summarized discussion of the remaining data sets}
}

So far, we have discussed the classification results for the various versions of
the third data set under a reduction via PCA. For the first and second data
sets, cf.\ Secs.~\ref{subsec:data:linsplit} and~\ref{subsec:data:mnist}, as well
as the third one with CAE reduction, we summarize their main results in
Fig.~\ref{fig:results_joint}~(a)-(d). The various panels show exclusively the
maximally achieved classification accuracies, from among the $50$ randomly
generated classical and quantum neural networks, obtained on the validation data
set after training on the training data set. We deem this the most crucial
performance indicator for comparing classical and quantum models.

For the third data set with CAE as the dimensionality reduction method, we give
the results in Fig.~\ref{fig:results_joint}~(d). In fact, they look very similar
to those presented in Fig.~\ref{fig:results_dataset_3_pca} in almost every
aspect. Most prominently, they also confirm that the best classical and quantum
models perform equally well despite the overall difficulty to classify the
strongly dimensionality-reduced industrial images. All classical and quantum
model performances lie furthermore within one standard deviation of each other,
thereby statistically confirming that none outperforms the other.

\change{%
In the following, we discuss the results of the simpler first and second data sets. We are aware that classical linear classifiers should work well for these simple problems and the usage of classical neural networks seems unnecessary. However, since we require classical models for which we can argue a similar complexity to the quantum models analyzed, and in order to ease direct comparison, we opted for classical neural networks even for these easier tasks.
}

In Fig.~\ref{fig:results_joint}~(a), the results for the first data set, i.e.,
the linear split of the hypercube, are shown. While each version of the data
set, distinguished by the number of data points $N$, achieves almost $100\%$
classification accuracies for classical models and small hypercube dimensions
$d$, their accuracies fall for larger $d$. This effect is even stronger for data
sets with fewer data points, which is explainable given the hypercube's growing
size but constant number of sampling points from which to learn the function
that splits it. In contrast, the best quantum models achieve classification
accuracies of $\sim 90\%$ at maximum while their average and lowest performances
(data not shown) show inferior performance compared to classical models as well.
Moreover, not even the best models' standard deviation reaches that of the
classical models. This observation of inferior performance is in agreement with
Ref.~\cite{Bowles2024}, which observed that a linear split through a hypercube
seems to be surprisingly difficult to classify for quantum models. Nevertheless,
by confirming former observations, we consider this a validation of our
methodology.

In Fig.~\ref{fig:results_joint}~(b), the results for the images of handwritten
digits, dimensionality-reduced by a CAE, are shown. While each version of the
data set, distinguished by their reduced dimension $d$ and the number of data
points $N$, achieves at least a classification accuracy of $>95\%$ for both
classical and quantum models, the accuracies for the quantum models are still
slightly lower. This becomes even more pronounced when using a PCA instead of
a CAE for dimensionality reduction with the corresponding results shown in
Fig.~\ref{fig:results_joint}~(c), although their standard deviations stay in
touch with those of the classical models. However, given the simplicity of the
classification task that originates from the simplicity of the corresponding
dimensionality reduced data, cf.\ Fig.~\ref{fig:overview}~(a) and~(b), it is not
surprising that both classical and quantum models solve the task well. This is
in line with other studies targeting the same data set and which report similar
classification accuracies~\cite{PRA.101.062327, Senokosov2024}. This serves as
another validation of our methodology.

\begin{figure}
  \centering
  \includegraphics{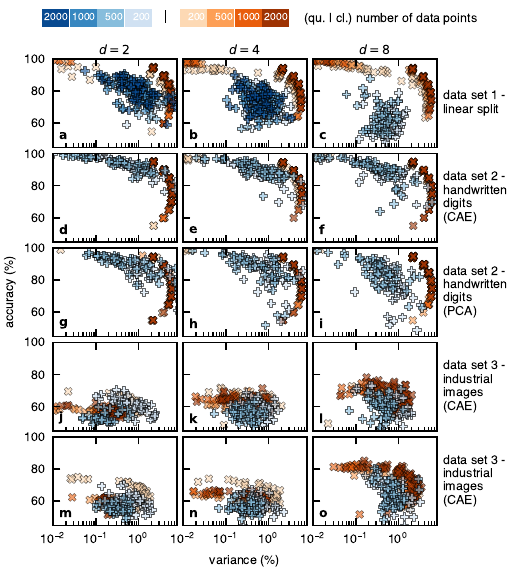}    
  \caption{%
    Overview of the variances as shown in Fig.~\ref{fig:results_dataset_3_pca}~(c)-(h) but for all data sets and reduction methods as indicated on the right-hand side of each row. We only show it for those data dimensions $d$ for which we have results for both classical and quantum neural networks.   
  }
  \label{fig:variances}
\end{figure}

\change{%
\subsubsection{Examination of learned functions and variances}
}

At last, we want to exemplarily compare the learned functions of the best
performing classical and quantum neural networks for all three data sets and
both reduction methods to $d = 2$, which is unfortunately the only dimension for
which we can visualize them. The results are shown in
Fig.~\ref{fig:overview}~(f)-(j) and~(k)-(o) for classical and quantum models,
respectively, and the corresponding best performing model architectures are
shown directly below or above the panels, respectively. We observe across all
data sets that the quantum models tend to learn more complex functions than the
classical ones despite some of the data's simplicity, especially for the first
and second data sets. Note that the classically learned functions do not change
qualitatively or in simplicity even if replacing the linear activation function
$\mathrm{ReLU}(x) = \max(0,x)$ with a non-linear one like $\tanh(x)$ (data not
shown). Following from Ref.~\cite{Pointing2024}, this may help evaluate if the
considered quantum neural networks have a poor inductive bias. Moreover, we also
want to stress that we observe the same behavior of the variances as discussed
in Fig.~\ref{fig:results_dataset_3_pca}~(c)-(h). We provide an overview of the variances with respect to the other data sets and reduction methods in Fig.~\ref{fig:variances}. In fact, the effect of the
quantum models generally having lower variance with respect to the starting
parameters is even stronger for the first and second data sets. Hence, this observation holds for all examined data sets.

\change{%
\subsubsection{Comparison with literature quantum models}
}
\change{%
As a means to compare our randomized quantum neural networks to known quantum models, we additionally benchmark the performance of three literature models, namely Chebyshev~\cite{Kreplin2024}, Hubregtsen~\cite{PRA.106.042431} and HighDim~\cite{Peters2021}.
To this end, we apply them to each data set and evaluate their performance as the averaged classification accuracy obtained from training with $10$ sets of initial parameters.
The comparison between our randomized quantum neural networks and these models (data not shown) yields that our best randomized models perform on par or even better than these specific literature models, with Hubregsten showing the best performance. This holds for all data sets.
}

\subsection{Analysis of correlations between performance and hyperparameters}
\label{subsec:results:rnd:analysis}

\begin{figure*}
  \centering
  \includegraphics{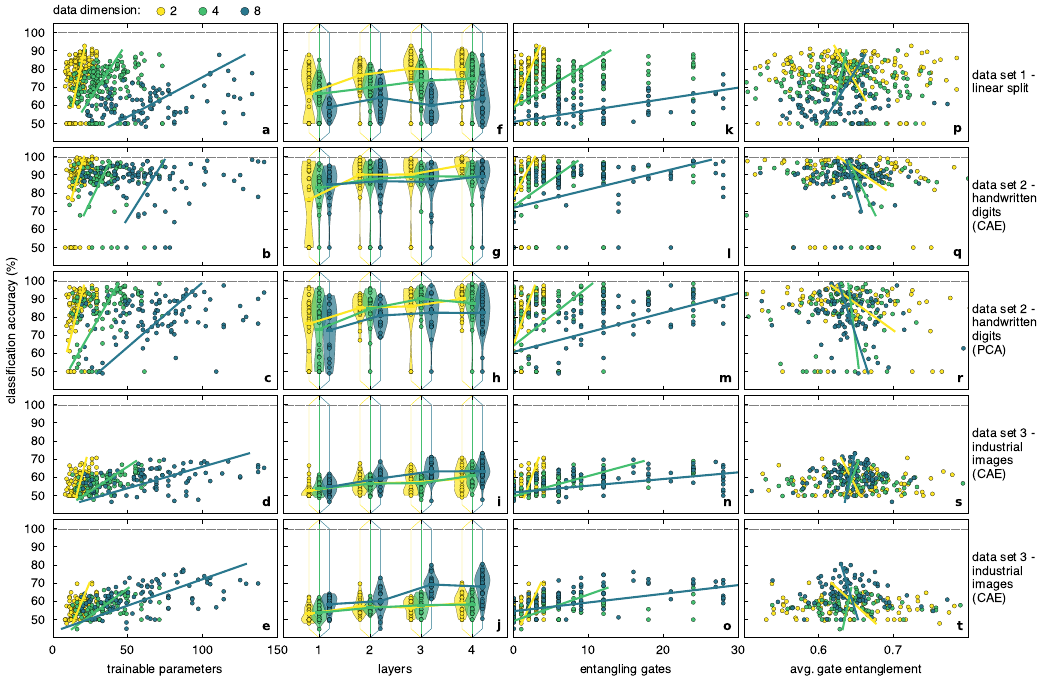}
  \caption{%
    Correlation analysis for the $50$ random quantum neural networks in terms of
    correlations between classification accuracy and various properties of the
    parametrized quantum circuit. The five rows correspond to the results
    obtained for the five data sets as indicated on the right. For simplicity
    and in order to have more data points for each data set, we merge all
    versions of each data set. To be precise, this implies that the presented
    data for the first data set contains the results gathered independently from
    training results with $N \in \{200, 500, 1000, 2000\}$ data points. For the
    second and third data set, it implies joining results gathered independently
    by training them with $N \in \{200, 500\}$. The colors distinguish the data
    dimensions. In panels~(a)-(e),the correlation between the joint number of
    trainable parameters in the PQC $\op{U}_{\rmpqc}(\myvec{x}_{i},
    \myvec{\phi}, \myvec{\theta})$, cf.\ Eqs.~\eqref{eq:pqc:full}
    and~\eqref{eq:pqc:red}, as well as its measurement operator
    $\op{O}(\myvec{\omega})$, cf.\ Eqs.~\eqref{eq:Opauli} and~\eqref{eq:Oising},
    and the classification accuracy is analyzed. In panels~(f)-(j), the
    correlation between the number $n_{\rmlayer, \rmq}$ of layers in the PQC and
    the classification accuracy is analyzed. While panels~(k)-(o) analyze the
    correlation between the number $N_{2\rmq}$ of entangling gates and the
    classification accuracy, panels~(p)-(t) analyze how their averaged gate
    entangling power, measured by Eq.~\eqref{eq:avgC}, for the trained model is
    correlated with the accuracy. The linear lines in every panel --- except panels~(f)-(j) --- solely indicate the general trend of the corresponding data and should not be taken as a fit to the data.
    Note that we do not take models yielding exactly $50\%$ classification
    accuracy after training into account for calculating the trend lines as
    those models are intrinsically untrainable, due to bad luck during
    randomization, and would otherwise impact the trend for actually trainable
    models.
  }
  \label{fig:corr}
\end{figure*}

\xchange{While Sec.~\ref{subsec:results:rnd:lit} shows that our best randomly generated quantum neural networks achieve similar performances to some tested literature models,}
We now analyze whether there are some common features which distinguish
well and badly performing random models. To this end, we first investigate the
correlation between the achieved classification accuracies and the total number
of trainable parameters in the model, since the latter is often connected to
a model's expressivity and thus performance. Figure~\ref{fig:corr}~(a) analyzes
this correlation for the first data set and reveals that the classification
accuracy of the hypercube data for $d = 2$ does not depend much on the number of
trainable parameters, as low and high accuracies can be observed for both low
and high numbers of trainable parameters. This changes for higher dimensional
hypercubes, which show an increasing correlation between accuracy and the number
of trainable parameters when moving to $d = 4$ and then further to $d = 8$. This
behavior is even stronger for the two versions of the third data set of
industrial images, distinguished by the employed reduction method and shown in
Fig.~\ref{fig:corr}~(d) and~(e). A possible explanation would be that due to the
increased complexity of the data, cf.\ Fig.~\ref{fig:overview}~(d) and~(e), it
takes more parameters to express a sufficiently complex function that is able to
at least partially approximate the exact but unknown classification function
$f: \mathbb{R}^{d} \rightarrow \{0, 1\}$. This is also in line with the results
for the second data set presented in Fig.~\ref{fig:corr}~(b) and~(c), where the
number of trainable parameters seems to play the least important role compared
to the other two data sets. This is explainable by the underlying data's low
complexity, cf.\ Fig.~\ref{fig:overview}~(a) and~(b). Interestingly, we find
that the correlations in Fig.~\ref{fig:corr}~(a), (d) and~(e) originate mainly
from correlations between the classification accuracy and the number of
trainable parameters in the PQC $\op{U}_{\rmpqc}(\myvec{x}_{i}, \myvec{\phi},
\myvec{\theta})$ and not so much due to correlations with the number of
trainable parameters in the measurement operator $\op{O}(\myvec{\omega})$. If
explicitly calculating the correlations statistics with respect to either of the
two (data not shown), we find very similar statistics to that shown in
Fig.~\ref{fig:corr}~(a)-(e) for the PQC and observe no correlations for the
measurement operator.

Since the total number of trainable parameters is intimately connected to the
number $n_{\rmlayer, \rmq}$ of layers in $\op{U}_{\rmpqc}(\myvec{x}_{i},
\myvec{\phi}, \myvec{\theta})$, we analyze its correlation with the
classification accuracies in Fig.~\ref{fig:corr}~(f)-(j). Interestingly, its
correlation is not as strong as expected for none of the data sets, despite
their observed correlation with the number of trainable parameters in
Fig.~\ref{fig:corr}~(a), (d) and~(e). We do, however, observe an increase in the
average classification accuracies across all data sets and all data dimension
when using two layers instead of one. For $n_{\rmlayer, \rmq} > 2$, the results
are again rather mixed with the classification accuracies stagnating for the
first and second data set. For the third data set, we still see a slight
increase in classification accuracies up to $n_{\rmlayer, \rmq} = 4$. However,
the general observation that $n_{\rmlayer, \rmq}$ does not seem to be crucial
when predicting a randomized quantum neural network's performance is also in
line with similar results observed in \change{our brief comparison with literature models}
\xchange{Fig.~\ref{fig:modelnames} for the literature models.}

Next, we examine the correlation between the classification accuracy and the
number of entangling gates in the PQC --- if the model's randomized generation
adds any in the first place. While Fig.~\ref{fig:corr}~(k)-(o) seem to indicate
a substantial correlation between the two quantities, we also do not want to
over-interpret the results. In our randomized models, each entangling gate comes
with a trainable parameter. Hence, having entangling gates or not heavily
impacts the number of trainable parameters and the observed correlation might
therefore be a reflection of Fig.~\ref{fig:corr}~(a)-(e) or vice versa.
A potentially more insightful quantity to calculate when analyzing the necessity
or impact of entangling gates, respectively entanglement, is shown in
Fig.~\ref{fig:corr}~(p)-(t). There, we calculate the entangling power of all
entangling gates using the gate concurrence~\cite{PhysRevA.63.062309} for every
gate's final training parameter. It is a measure $\mathcal{C}(\op{U}) \in [0,1]$
for all two-qubit gates $\op{U} \in \mathrm{U}(4)$ and attains $0$ ($1$) if the
corresponding gate $\op{U}$ is non-entangling (maximal entangling). We do not
find a correlation between a model's classification accuracy and the model's
averaged gate entangling power
\begin{align} \label{eq:avgC}
  \bar{\mathcal{C}}
  =
  \frac{1}{N_{2\rmq}} \sum_{i=1}^{N_{2\rmq}}
  \mathcal{C}\left(\op{U}_{i, \mathrm{trained}}\right),
\end{align}
with $i$ running over all $N_{2\rmq}$ trained entangling gates $\op{U}_{i,
\mathrm{trained}} \in \mathrm{U}(4)$ present in the model's PQC. The same holds
if analyzing the averaged change in gate entanglement power (data not shown),
which is given by
\begin{align} \label{eq:avgCchange}
  \bar{\mathcal{C}}_{\mathrm{change}}
  =
  \frac{1}{N_{2\rmq}} \sum_{i=1}^{N_{2\rmq}} \left[
    \mathcal{C}\left(\op{U}_{i, \mathrm{trained}}\right)
    -
    \mathcal{C}\left(\op{U}_{i, \rminit}\right)
  \right],
\end{align}
with $\op{U}_{i, \rminit}$ the untrained version of $\op{U}_{i,
\mathrm{trained}}$ determined by the initial training parameters. This analysis
reveals that high classification accuracies are neither correlated with larger
gate entanglement powers nor with the optimizer increasing it during training.
\change{%
This highlights the highly non-trivial effect that entanglement has on a model's performance. While having entangling gates seems to be mostly beneficial, they are not necessarily required for their entanglement-generating properties.
}

Besides the correlations analyzed in Fig.~\ref{fig:corr}, we also analyze the
correlations between the classification accuracy and various other quantities
(data not shown), for instance with respect to the number of qubits $n_{\rmq}$,
the form of the measurement operator $\op{O}(\myvec{\omega})$, the form of the
entangling structure or the employed feature encoding. We did not find any
correlations for any of these quantities for any of the data sets. We therefore
conclude that our correlation analysis in general does not yield any clear
indicator or rule on how most of the varied hyperparameters of the random
quantum neural networks should be chosen for ideal performance. The only
deducible rule seems to be that having more trainable parameters in the quantum
models is almost always beneficial at the tested problem scale, and irrelevant at worst.

However, we may deduce something when analyzing only the best-performing quantum models. Upon inspecting in detail the structure of the three best-performing quantum models for all combinations of data set, dimensionality reduction method and number of features, we report that only five out of these $90$ models were circuits without any entangling gates. Moreover, analyzing the cases where we have $d = 8$ features per data point, only three out of the $30$ best models did not use all-to-all entanglement. For these specific cases in particular, the best-performing quantum model always used all-to-all entanglement.

\subsection{Comparison of quantum model performances across data sets}

\begin{figure}
  \centering
  \includegraphics{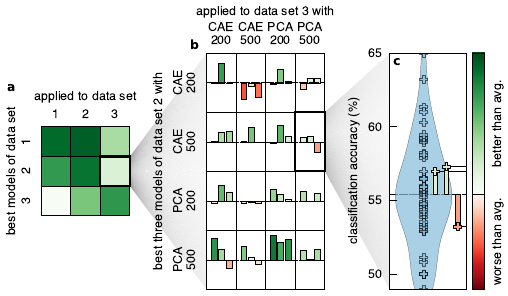}
  \caption{%
    Cross-comparison of quantum neural network performances between different
    data sets. Panel (a) shows the color-coded averaged model performance when
    applying the best three quantum models of each data set to itself and each
    of the other data sets. Panel (b) exemplifies the averaging process by
    showcasing the individual results that are being averaged and shown in panel
    (a). In detail, panel (b) shows the results when taking the three best
    performing models from each version of the second data set, i.e., either $N
    = 200$ or $N = 500$ data points and either CAE or PCA as dimensionality
    reduction method, and applying each of them to each version of the third
    data set, i.e., again using either $N = 200$ or $N = 500$ data points and
    either CAE or PCA as dimensionality reduction method. Panel (c) illustrates
    the color code used in panels (a) and (b) exemplarily. The greener [redder]
    the tile in panel (a) or bars in panel (b), the higher above [lower below]
    the three best models from the original data set (data set 2, CAE reduction,
    $500$ data points in panel (c)) perform with respect to the average random
    model's performance on the new data set (data set 3, PCA reduction, $500$
    data points in panel (c)). This is also illustrated by the non-linear
    colorbar on the right of panel (c), whose maximum/minimum/center is aligned
    with the best/worst/average performances given by the $50$ random quantum
    neural networks generated for the new data set (data set 3, PCA reduction,
    $500$ data points). All data sets have dimensionality $d = 4$.
  }
  \label{fig:cross}
\end{figure}

We now utilize the fact that we have classification results for various randomly
generated quantum neural networks on various data sets. This allows a cross-data
set comparison of model performances in the sense that we can compare whether
models which perform well on one particular data set also perform well on other
data sets. This way we may assess whether some of the randomly generated models
seem to be well-performing in general. To this end, Fig.~\ref{fig:cross}~(a)
shows the cross-data set performances of the best performing models from one
data set when applied to any of the other data sets. However, in order to have
enough statistical data to average over, we always take the best three models
--- quantified by the three highest classification accuracies --- and train them
on other data sets. Moreover, in Fig.~\ref{fig:cross}~(a), we also average over
the different versions of each data set, i.e., we average over the results when
applying the best three models for one particular version of a data set to one
particular version of another data set. The different versions are distinguished
by their number of data points or the employed reduction method in the case of the
second and third data sets. The data dimensionality is kept identical to allow
a model's applicability to a different data set. It is exclusively $d = 4$ for
Fig.~\ref{fig:cross}. Figure~\ref{fig:cross}~(b) shows in detail how the results
of Fig.~\ref{fig:cross}~(a) are obtained for the example of applying the best
models of the second data set to the third data set. The various panels in
Fig.~\ref{fig:cross}~(b) show the results when using the best three models for
each of the four versions of the second data set, distinguished by $N \in \{200,
500\}$ and CAE or PCA as reduction method, to the same four versions but of the
third data set. For one of the panels in Fig.~\ref{fig:cross}~(b), we show how
its cross-data set performance is calculated in Fig.~\ref{fig:cross}~(c). In the
chosen example, this implies how the best three models for the second data set
with $N = 500$ data points and CAE as reduction method (called original data set
in the following) perform when trained for the third data set with $N = 500$
data points and a PCA reduction (called new data set in the following). The
three model performances --- evaluated again by averaging each model's results
for $10$ set of random initial training parameters --- are then compared to the
best, average and worst performances of the $50$ random quantum models
previously generated for the new data set. A model from the original data set is
considered good (bad) if it performs better (worse) than the average performance
of the new data set. This is indicated with green (red) bars in
Fig.~\ref{fig:cross}~(c). The colors of these bars carry the relevant
information in Fig.~\ref{fig:cross}~(b) and their average determines the results
presented in Fig.~\ref{fig:cross}~(a).

With this explanation in mind, we may discuss Fig.~\ref{fig:cross}~(a) and first observe that every model which performs well on any of the three data sets also performs well on the second data set, which we attribute to the latter data set's simplicity. The contrary, however, is not true. The models which perform well on the first and second data sets seem to perform rather mediocrely on the third one, and the best performing models on the third data set perform likewise mediocrely on the first data set.
This separation seems to be correlated with the data's complexity, cf.\ Fig.~\ref{fig:overview}~(a)-(e), specifically with how separable the two classes are. While the first and second data sets are rather simple, in the sense that they are almost always linearly separable, the third data set is more complex and the classes are never linearly separable.
We thus conclude that, while our randomized generation of quantum neural networks seems to yield decent models which perform mostly well, and at worst mediocrely on all tested data sets, in order to have the best possible performance, customized models need to be crafted in a data-set-dependent way.

\section{Results from convolutional neural networks}
\label{sec:results:conv}
After the discussion of the classification results of the dimensionality reduced
data sets from Sec.~\ref{sec:data} in the previous Sec.~\ref{sec:results:rnd}
employing the first methodology, cf.\ Sec.~\ref{subsec:method:rnd}, we now turn
towards presenting the results employing the second methodology, cf.\
Sec.~\ref{subsec:method:conv}, to classify the original, non-reduced image data.

\subsection{Comparison of classical and hybrid convolutional neural network
performances}
\label{subsec:results:conv:CvsH}

Similar to the previous analysis of the dimensionality reduced data sets, we
start our analysis for the non-reduced data sets by comparing the performance in
terms of classification accuracies for the fully-classical and quantum-classical
hybrid scheme --- here given by CNNs and QCCNNs, respectively. However, while
running the first simulations for the handwritten-digit data sets with $N = 200$
and $N = 500$ images, we noticed a large discrepancy between the classification
accuracy of CNNs and QCCNNs for the $N = 200$ data set, in favor of the latter
models. This discrepancy could not be observed for the data set with $N=500$. To
study this possible advantage in few-data scenarios in more depth, we decided to
generate more data sets with various $N$, following the same generating
principles as described in Sec.~\ref{sec:data}. In the following, we apply our
(quantum) convolutional models to data sets with size $N \in \{50, 100, 200,
300, 400, 500\}$.

\begin{figure}
  \centering
  \includegraphics{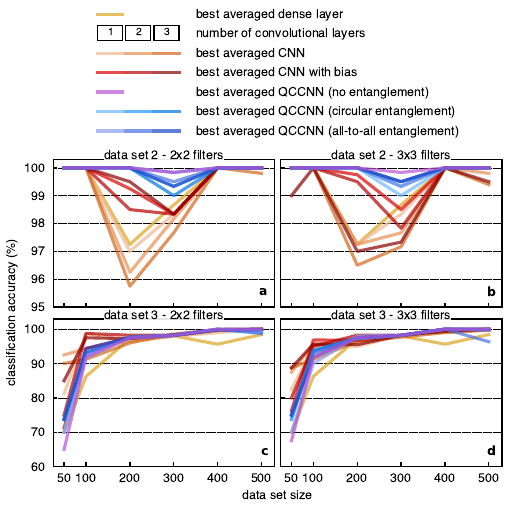}
  \caption{%
    Performance comparison between fully-classical (warm colors) and
    quantum-classical hybrid (cool colors) convolutional neural networks (CNN
    and QCCNN, respectively) in terms of the classification accuracies of each
    tested model as function of data set size $N$, for both non-reduced image
    data sets, cf.\ Secs.~\ref{subsec:data:mnist} and~\ref{subsec:data:trumpf}.
    A quantum model is determined by a particular choice of three
    hyperparameters: size of the quantum-convolution filter, entanglement
    structure in the quantum circuit used in the quantum convolution, and number
    of layers in this quantum circuit. The performance of a single dense
    classical layer with two neurons and softmax activation function is given as
    baseline (golden line). The data is also separated by the size of the
    (quantum) convolution filter such as to not overcrowd each plot.
  }
  \label{fig:qccnn_acc_vs_dataset_size}
\end{figure}

In Fig.~\ref{fig:qccnn_acc_vs_dataset_size}, we summarize the results obtained
with all tested models, in terms of classification accuracies on the validation
data sets as a function of the data set size $N$. The handwritten-digit data
sets are shown in panels~(a) and~(b) and the industrial-image data sets in
panels~(c) and~(d). As before, the classical models are depicted as warm-color
lines, while the hybrid models as cool-color lines.

\change{%
\subsubsection{Discussion of data set 2 --- handwritten digits}
}
For the handwritten-digit
data sets, QCCNNs consistently reach classification accuracies close to $100\%$,
while CNNs achieve as low as $96\%$ for some data set sizes. Interestingly,
QCCNNs without entangling gates show the best performance (purple lines), which
indicates that this advantage is most likely not due to inherently quantum
mechanical properties, but probably due to the algorithmic structure, which
might even be efficiently classically simulable. Moreover, the initial
assumption of quantum advantage in few data scenarios is not confirmed. In fact,
for the smallest data set size, all classical and hybrid models achieve $100\%$
accuracy. This confirms that, in this regime of very few data, the choice of
individual images has a large impact in defining the problem hardness.

\change{%
\subsubsection{Discussion of data set 3 --- industrial images}
}
For the industrial-image data sets, the performances of fully-classical and
quantum-classical hybrid models are very similar, with the former showing better
generalization ability for the smallest-sized data sets of $N = 50$ and $N
= 100$ images. For these data sets, QCCNNs with entangling gates perform
slightly better than those without. However, for the larger-sized data sets,
both CNNs and QCCNNs, as well as all tested varieties of the latter, perform
almost identically. We highlight that, even for the most challenging data set,
QCCNNs show averaged accuracies below $95\%$ only for data sets with $N \in
\{50, 100\}$. On the one hand, this indicates that these models are already
expressive enough to deal with some industry-relevant use cases. On the other
hand, the considered industrial images may still not exhibit enough complexity
to fully benchmark these models, thus preventing us from drawing more rigorous
conclusions.

\vspace{1em}

For both data sets, similar arguments can be made when analyzing the best
averaged model's cross entropy loss values on the validation data sets (data not
shown). We note that, despite QCCNNs without entangling gates achieving
a similar performance to QCCNNs with entangling gates, the former took more
iterations to converge both in classification accuracy and cross-entropy loss in
most scenarios (data not shown). However, this effect was more apparent for the
scenarios with smaller-sized data sets. We further highlight that, for the
handwritten-digit data sets, while most classical and hybrid models converged in
validation accuracy within the $100$ epochs, some did not. This encompasses in
particular the QCCNN with a $3 \times 3$ quantum convolution filter and without
entangling gates for $N = 300$, as well as multiple CNNs for $N = 200$. On the
other hand, for the industrial-image data sets, some models did not converge
within the $100$ epochs, specially for the smaller data sets with $N\in\{50,
100\}$, indicating that higher accuracies than the ones displayed may in general
be achievable with these models.

\begin{figure}
  \centering
  \includegraphics{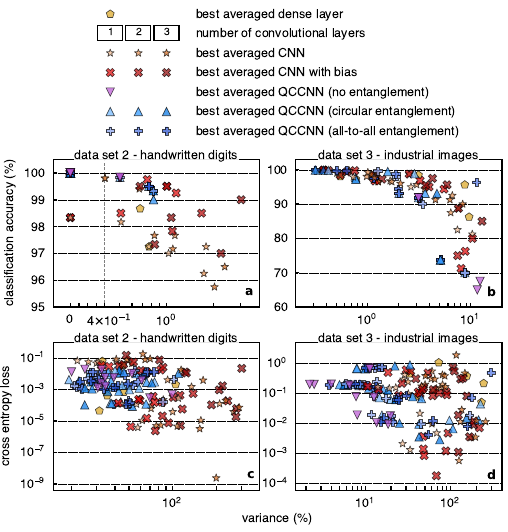}
  \caption{%
    Correlation analysis for the CNNs (warm colors) and QCCNNs (cool colors) in
    terms of correlations between classification accuracy [cross entropy loss]
    on the validation data sets and its variance, for both image data sets. The
    accuracy [loss] analysis can be seen in panels~(a) and~(b) [(c) and~(d)].
    For simplicity and to have more points for each data set, each panel
    contains results for models with all sizes of (quantum) convolution filter,
    each trained with $N \in \{50, 100, 200, 300, 400, 500\}$ data points. The
    performance of a single dense classical layer with two neurons and softmax
    activation function is also analyzed for comparison (golden pentagons). In
    order to represent points with null variance, the variance axis in panel~(a)
    has a symmetrical logarithmic scale where the range $\left[0, 4 \times
    10^{-1}\right]$ is linear and $\left[4 \times 10^{-1}, \infty\right)$ is
    logarithmic. A vertical dashed line separates these two ranges for clarity.
  }
  \label{fig:qccnn_acc_loss_vs_variance}
\end{figure}

\change{%
\subsubsection{Examination of performance variances}
}
Figure~\ref{fig:qccnn_acc_loss_vs_variance}~(a) and~(b) show the relationship
between the models' performance metrics and their variance across the $10$ runs
with different initial parameter sets, for both image data sets. The same color
structure as in Fig.~\ref{fig:qccnn_acc_vs_dataset_size} encodes the results.
For simplicity and to maximize the number of points available for the analysis,
each panel contains results for models with all sizes of (quantum) convolution
filters, trained with each data set size. On average, QCCNNs show lower
performance variance than CNNs, which agrees with our observation for random
neural networks, cf.\ Fig.~\ref{fig:results_dataset_3_pca}~(c)-(h). This lower
variance of models with quantum-circuit components can be seen even clearer when
analyzing the relationship between the classification accuracies' variance and
the underlying cross-entropy loss shown in
Fig.~\ref{fig:qccnn_acc_loss_vs_variance}~(c) and~(d) for each data set.

We observe a negative correlation between classification accuracy and its
variance, cf.\ Fig.~\ref{fig:qccnn_acc_loss_vs_variance}~(a) and~(b), which may
be interpreted as better-performing models having a loss landscape whose
better-performing minima are more easily reachable independently of the initial
parameters. In particular, all models achieving an accuracy of $100\%$ for the
handwritten-digit data sets show no accuracy variance. Thus, their performance
was completely independent of the chosen initial parameters.

\subsection{Analysis of correlations between performance and hyperparameters}

\begin{figure*}
  \centering
  \includegraphics{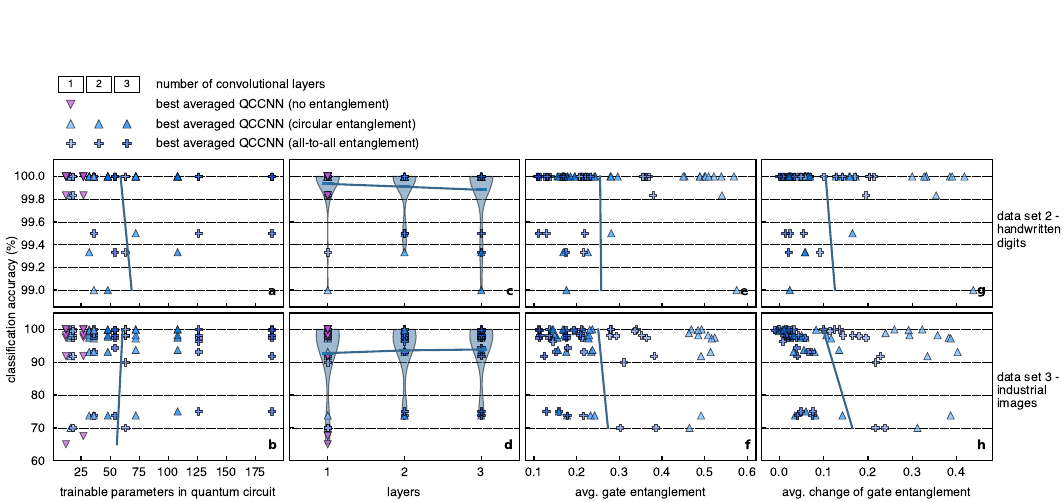}
  \caption{%
    Correlation analysis for the quantum-classical hybrid convolutional neural
    networks (QCCNN) in terms of correlations between classification accuracies
    and various properties of the parametrized quantum circuit used for the
    quantum convolution. Each row corresponds to the results obtained for the
    data set indicated on the right. For simplicity and to have more points for
    each data set, we merge all QCCNNs and data set sizes used. Therefore, each
    panel contains results for models with all sizes of quantum-convolution
    filter, all entanglement structures of the circuits used for its quantum
    convolution, and number of layers in this quantum circuit, each trained with
    $N \in \{50, 100, 200, 300, 400, 500\}$ data points. Visually, we still
    separate the latter two hyperparameters (entanglement structure and number
    of layers) by color. In panels~(a) and~(b), we analyze the correlation
    between classification accuracy and number of trainable parameters in the
    quantum circuit used in the quantum convolution. In panels~(c) and~(d), we
    analyze the correlation between classification accuracy and the number of
    layers in the quantum circuit. In panels~(e) and~(f), we analyze the
    correlation between classification accuracy and averaged gate entangling
    power $\bar{\mathcal{C}}$, cf.\ Eq.~\eqref{eq:avgC}, of the models after
    training. Finally, panels~(g) and~(h) analyze the correlation between
    classification accuracy and the change of this average gate entanglement
    power before and after training, cf.\ Eq.~\eqref{eq:avgCchange}. The QCCNNs
    without entangling gates are omitted in panels~~(e)-(h). The linear lines in
    every panel --- except panels~(c) and~(d) --- solely show the general trend of all
    the data and should not be taken as a fit to the data. In panels~(c) and~(d), the lines connect the average accuracy of
    each set of QCCNNs with a specific number of layers in its quantum circuit.
   }
  \label{fig:qccnn_correlation_analysis}
\end{figure*}

We now perform the same analysis we did for the random quantum neural networks,
cf.\ Sec.~\ref{subsec:results:rnd:analysis}, in terms of classification accuracy
and various properties of the quantum circuits used in the QCCNNs, in order to
identify key ingredients of either well or badly performing models.
Figure~\ref{fig:qccnn_correlation_analysis} shows some of the analyzed
properties of the quantum circuits of all QCCNNs used, for each data set and
each trained with all data set sizes available, in order to maximize the number
of points available for the analysis. Visually, we opt to separate by color the
entanglement structures of the circuit used for the quantum convolution and the
number of layers in this quantum circuit. However, we consider all data points
when looking for any correlations or interesting patterns. Since, for each data
set type, the number of trainable parameters in the final dense classical layer
of each QCCNN is similar, changing only if the images need padding before the
quantum convolution, we opt to analyze the correlation between the achieved
classification accuracy and the number of trainable parameters in the quantum
circuit of the quantum convolution. This is represented in
Fig.~\ref{fig:qccnn_correlation_analysis}~(a) and~(b), for the data sets
indicated on the right. The accuracy does not seem to depend on the number of
trainable parameters in the quantum circuits, since there are both well and
badly performing models with both few and many trainable parameters in the
quantum layer. However, the amount of data is too limited to draw
a statistically significant conclusion from this.

Looking at Fig.~\ref{fig:qccnn_correlation_analysis}~(c) and~(d), which show the
relationship between accuracy and the number of layers in the quantum circuit of
the QCCNN, we also find no clear correlations. This is in line with the
observations made for random quantum neural networks, cf.\
Fig.~\ref{fig:corr}~(f)-(j). It is again hard to extract more rigorous
conclusions, especially because the median accuracy for each set of QCCNNs with
a specific number of layers in its quantum circuit is always very close to
$100\%$.

\begin{figure}
  \centering
  \includegraphics{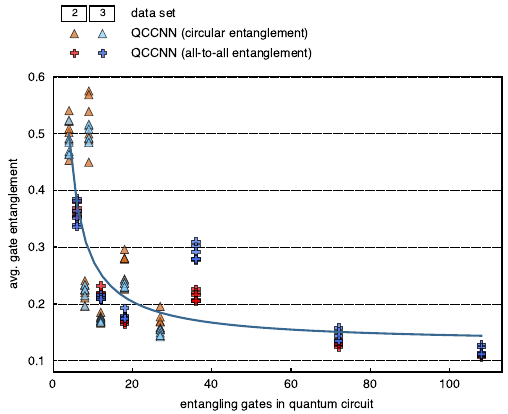}
  \caption{%
    Averaged gate entangling power $\bar{\mathcal{C}}$, cf.\
    Eq.~\eqref{eq:avgC}, of QCCNNs with entangling gates, after training, as
    a function of the number $N_{2\rmq}$ of entangling gates in the quantum
    circuit used for the quantum convolution. The data points of both
    Fig.~\ref{fig:qccnn_correlation_analysis}~(e) and~(f) are shown here. Data
    points of QCCNNs trained with the handwritten-digit data sets have warm
    colors, while those trained with the industrial-image data sets have cool
    colors. QCCNNs with circular entanglement structure have lighter tones,
    while those with all-to-all entanglement have darker tones. The hyperbolic
    curve was fit to all the data points.
  }
  \label{fig:qccnn_avg_gate_ent_vs_n_entangling_gates}
\end{figure}

Since the entangling gate structures were fixed and all entangling gates
parametrized, the correlation between the number of trainable parameters and the number
of entangling gates is very high, by definition. Therefore, we opt not to show this correlation
plot. To analyze the usefulness of entanglement in QCCNN models, we calculate
the averaged gate entangling power $\bar{\mathcal{C}}$, cf.\
Eq.~\eqref{eq:avgC}, of the models after training, as done in
Sec.~\ref{subsec:results:rnd:analysis}. For this analysis, we exclude QCCNNs
without entangling gates for obvious reasons. The relationship between this
metric and classification accuracy is shown in
Fig.~\ref{fig:qccnn_correlation_analysis}~(e) and~(f), for each data set. We do
not find a correlation between the classification accuracy of a QCCNN and its
averaged gate entangling power $\bar{\mathcal{C}}$. However, for the
handwritten-digit data set, we observe that most models with larger
$\bar{\mathcal{C}}$ reach a classification accuracy of $\sim 100\%$ on the
validation data set. Conversely, if a model performs badly, it is most likely one
with lower $\bar{\mathcal{C}}$. Nevertheless, the amount of data is again too
limited to make a statistically significant statement. We still note that QCCNNs
with more entangling gates show a lower $\bar{\mathcal{C}}$, which we further
analyze in Fig.~\ref{fig:qccnn_avg_gate_ent_vs_n_entangling_gates}. In fact,
these two quantities seem to be inversely proportional. Fitting the expression
\begin{align}
  \bar{\mathcal{C}}\left(N_{2\rmq}\right)
  =
  \frac{a}{N_{2\rmq}} + b,
\end{align}
with $N_{2\rmq}$ the number of entangling gates in the quantum circuit used for the quantum convolution, to all data points, we get the curve depicted in Fig.~\ref{fig:qccnn_avg_gate_ent_vs_n_entangling_gates}, with parameters $a = 1.47 \pm 0.10$ and $b = 0.13 \pm 0.01$. We may explain this inverse relation as a consequence of the total gate entangling power being distributed among all entangling gates for each model.

\change{%
This inverse relationship raises a fundamental question about what metrics really capture the utility of entanglement in quantum models. There is certainly nuance between: (i) the binary metric of presence or absence of entangling gates, (ii) the discrete metric of number of entangling gates, and (iii) the continuous metric of gate entangling power. This inverse proportionality between the metrics (ii) and (iii) challenges the intuitive assumption that more entangling gates likely lead to more useful entanglement for machine-learning tasks --- an assumption often made in the design of quantum circuits for machine learning.
}
While a deeper analysis is needed, this suggests that, for problems which require a specific amount of entanglement generated by a quantum model, models with \xchange{a small amount of entangling gates may be as performant as models with many such gates} \change{concentrated entanglement through fewer, more powerful gates may be as performant as models with entanglement distributed across many weaker gates. Thus, we suggest that practitioners should focus on optimizing gate entangling power rather than simply increasing the number of entangling gates in their studies}.
Moreover, training the former should be easier through classical simulation, since they can have many fewer parameters and, thus, require fewer circuit simulations to calculate gradients.

We perform the same analysis for the average change of gate entanglement power
$\bar{\mathcal{C}}_{\mathrm{change}}$, cf.\ Eq.~\eqref{eq:avgCchange}, during
training, shown in Fig.~\ref{fig:qccnn_correlation_analysis}~(g) and~(h), for
each data set. We also find no correlation between this metric and the
classification accuracy. However, we note that, for both data sets, roughly $2/3$
of the individual entangling gates across all QCCNNs increase their entangling
power during training. Moreover, virtually all models show a positive
$\bar{\mathcal{C}}_{\mathrm{change}}$, with only five models out of $144$
showing a slightly negative $\bar{\mathcal{C}}_{\mathrm{change}}$. While this
seems like a positive argument for the use of entanglement as a resource in
quantum machine learning, we must keep in mind that quantum-classical hybrid
models without entangling gates performed similarly well as the ones with entangling gates, in most scenarios.

\section{Summary and conclusions}
\label{sec:conclusion}

In this study, we conduct a systematic investigation of classical- and quantum-machine-learning approaches for binary image classification across data sets of increasing complexity. Our work provides a uniquely comprehensive hyperparameter optimization analysis --- rarely conducted in quantum-machine-learning literature --- offering insights into which quantum circuit properties actually contribute to model performance. We evaluate both classical and quantum-classical hybrid approaches using three progressively complex data sets: (i) artificial hypercube data, (ii) MNIST handwritten digits (0's and 1's)~\cite{deng2012mnist}, and (iii) \xchange{real-world} industrial images \change{of practical relevance} from laser cutting machines.

Our methodological framework employs two complementary approaches: (1) randomized (quantum) neural networks applied to dimensionality-reduced data, and (2) (hybrid) convolutional neural networks applied to full image data. Both methodologies effectively compare fully-classical versus quantum-classical hybrid pipelines. Dimensionality reduction through convolutional autoencoders (CAE) and principal component analysis (PCA) serves as a necessary preprocessing step for fully-quantum models, as the high dimensionality of industrial data sets would otherwise make them computationally intractable for quantum processing. The first methodology employs either classical or quantum neural network classifiers on dimensionality-reduced data, while the second uses either classical CNNs or quantum-classical hybrid convolutional neural networks (QCCNNs)~\cite{Henderson2019}.

Quantitatively, using classification accuracy on validation data as our performance metric, we observe that classical and quantum methods achieve statistically equivalent classification accuracies across most tested data sets. For the industrial image data, maximum averaged accuracies of quantum models reached approximately $80\%$ for dimensionality-reduced data with $d=8$ features and between $95\%$ and $100\%$ for full-image classification with convolutional approaches, except for the smallest data sets containing 50 and 100 total images. Our results from the two simple data sets, the linear split of a $d$-dimensional hypercube and the $0$'s and $1$'s from MNIST's handwritten images, are in good agreement with previous studies~\cite{Bowles2024}, confirming that MNIST is trivial for both classical and quantum models while, for the linearly-split hypercube, despite its simplicity, it is already surprisingly challenging to reach classification accuracies beyond $90\%$. However, these two simple data sets are merely used to benchmark our methodologies and validate their application for the industry data set. This empirical equivalence in performance highlights a significant finding: despite substantial theoretical advancement in quantum machine learning, practical implementation on industrial data sets shows no clear \xchange{quantum} \change{practical} advantage in classification accuracy. \change{From an industry perspective, we do not observe quantum models to offer any practical advantage for the considered binary classification tasks and classical data sets. The most likely reason as to why we did not observe a practical advantage, if there is one to observe in the first place, is that we needed to restrict our quantum models to small and medium sizes in order to keep them simulable. This might change with more powerful quantum hardware that supports executing quantum models which reduce or fully avoid any classical processing.}

An intriguing finding is that quantum and hybrid models demonstrate significantly lower variance in classification accuracies with respect to initial training parameters compared to classical approaches. This was consistently observed across all data sets and methodologies, suggesting improved stability in quantum models when optimal initialization parameters are unknown --- a potentially valuable property for practical applications \change{as it implies more predictable performance}.
Conversely, when analyzing learned decision boundaries, classical models consistently produced simpler functions while quantum models generated notably more complex classification boundaries, even for data sets with easily-separable classes.
This tendency towards \change{non-linear} complexity despite no performance advantage \xchange{might indicate a lack of appropriate inductive bias} \change{may indicate a lack of appropriate regularization mechanisms which encourage simpler solutions --- a simplicity bias, often beneficial for generalization ---} in the quantum models used~\cite{Pointing2024}.
\change{
The disconnect between our findings and approaches focused on inductive biases \cite{Bowles2023,Gili2024} suggests that practical advantage may require careful design of quantum architectures, such that they are able to fully exploit known structures in the data, rather than relying on generic quantum properties.
}

Our comprehensive hyperparameter analysis revealed surprisingly few consistent correlations between model-architecture choices and performance. Among all parameters analyzed (feature encoding methods, number of circuit layers, entangling gate type and structure, gate entangling power, measurement operators), only the number of trainable parameters showed a consistent positive correlation with model performance. However, this is not a general statement as, at some point, we expect to run into the overfitting regime. The inconsistent importance of hyperparameters across our experiments --- particularly regarding entanglement structure and gate entangling power --- illuminates the current limited theoretical understanding of how quantum models actually function and what determines their performance.

\xchange{Upon comparing our randomized quantum neural networks with some literature architectures, we observe that our best randomized quantum models can easily compete with the top performing literature model~\cite{PRA.106.042431} among the ones we tried, including Chebyshev~\cite{Kreplin2024}, Hubregtsen~\cite{PRA.106.042431}, and HighDim~\cite{Peters2021}.}
For the randomized quantum neural networks, we found no correlation between model performance and total averaged gate entanglement power or its change during training, with gate entangling power quantified using gate concurrence~\cite{PhysRevA.63.062309}. However, $\sim 94\%$ of the best performing models did feature entangling gates, often with all-to-all connectivity. Nevertheless, we note that this entanglement structure can always be mapped to a linear one with a polynomial gate overhead~\cite{PhysRevApplied.23.034022}. Thus, while the type of entangling structure may not be a crucial factor, the non-negligible entangling power of the top-performing models is a hopeful observation. \xchange{Further testing the usefulness of entanglement in quantum circuits with a fixed amount of trainable parameters is a goal for future work.}

For QCCNNs, models without entangling gates performed equally well as those with entanglement in most scenarios, though they required more iterations to converge in both classification accuracy and cross-entropy loss, especially with limited training data. This suggests that, while entanglement may not be a crucial factor for the performance of quantum-classical hybrid models, it may still be beneficial for their training.
\change{%
Nevertheless, we must note that the QCCNNs tested here have significantly more classical training parameters than quantum ones. Thus, their expressive power is likely dominated by their final classical dense layer. Hence, the expressive power of the quantum convolutional layer might be too limited, irrespective of it having entangling gates or not.
}
Interestingly, for the QCCNNs, we still observe a positive change in averaged gate entanglement power due to training in virtually all scenarios. Most importantly, we also observed an inverse relationship between the number of entangling gates and per-gate entanglement power, suggesting that the total entanglement capability is distributed among available gates --- a finding with implications for efficient quantum circuit design.

While a deeper statistical analysis would be needed to confirm the generality of these properties, \change{the entanglement findings in this study highlight a critical gap in our current understanding of quantum resource utilization. While we can measure various aspects of entanglement --- from gate-level concurrence~\cite{PhysRevA.63.062309} to circuit-wide entanglement structures --- the relationship between these metrics and actual computational utility remains poorly understood. The fact that models without entangling gates can perform comparable to those with entanglement in QCCNNs, while the opposite trend appears in the quantum neural networks of the first methodology, suggests that entanglement requirements may be context-dependent and influenced by other architecture factors, such as the relative importance of classical versus quantum processing components.
Given the availability of entanglement is one of the key resources separating quantum and classical methods, we strongly believe that improving the general understanding of the role of entanglement in quantum machine learning models is a crucial future research direction.
Since a major difference between the two methodologies is the number of trainable parameters, we believe a good starting point for future work would be to test the usefulness of entanglement in quantum circuits with a fixed number of trainable parameters.
Crucial but specific open questions are: (i)``which type of data structures benefit from entanglement?'', (ii) ``how does the amount of classical processing in a hybrid model impact the usefulness of entanglement?'' and (iii) ``how to best quantify the entanglement in a quantum circuit and how to relate it to model performance?''.
Regarding this last open question, many metrics can be considered, such as the structure of the entangling gates, their type, their number, and the per-gate or total entangling power.
Moreover, looking at state entanglement metrics could prove more insightful than gate-level metrics, since they not only take into account the circuit unitary, but also the data being processed.
This can be tackled by quantifying the full state entanglement as the initial (non-entangled) state propagates through the quantum circuit, giving rise to potentially complicated multi-partite entanglement before measurement. This can reveal whether the entangling capabilities of any gates are actually used to create entanglement in the states generated with each specific data set. Unfortunately, this analysis is computationally very expensive, which is why we did not include it in this study.
}

Our cross-data set analysis reveals that quantum-model transferability is limited, with performance possibly dependent on data-separability characteristics. Models performing well on linearly-separable data showed mediocre performance on the more complex, non-linearly separable industrial images. While there seems to be a correlation between model architecture and class separability, further research is needed to confirm this hypothesis. This suggests that, despite the universal approximation capabilities theoretically attributed to quantum models, custom architecture design remains necessary for optimal performance on specific data sets.

We approached this study with fairness considerations in both directions: restricting the parameter count of classical models to match quantum models, while limiting quantum model size to ensure \change{compatibility with near-term quantum devices and} classical simulability. Despite these constraints, our findings provide valuable insights into the current state of quantum machine learning for industrial image classification. The absence of performance advantage, combined with our systematic hyperparameter optimization analysis, illustrates the significant gap between theoretical quantum advantage propositions and practical implementations.

The question of whether one model outperforms another also strongly depends on the performance metric. Here, we exclusively use the classification accuracy on the validation data set, although other metrics such as convergence speed or training time could, in principle, be employed. However, for the latter, it would be difficult to make a tangible statement without running all quantum circuits on actual quantum hardware --- instead of simulating them --- and factoring in data loading and circuit execution times as well as their frequent repetition until enough statistics would be gathered. Hence, until sufficiently potent quantum hardware is available, it is difficult to judge the practical potential of quantum-machine-learning methods, since, up to then, either a classical preprocessing of data or a restriction of quantum resources is required --- both heavily impacting the conclusions one may draw from any results. Nevertheless, our comprehensive hyperparameter study provides a methodological framework and empirical baseline for future quantum-machine-learning research focused on realistic industrial applications, offering crucial insights into the current limitations and potential directions for advancement in this rapidly evolving field.

\begin{acknowledgments}
  This work was supported by the Federal Ministry for Economic Affairs and Climate Action (Bundesministerium für Wirtschaft und Klimaschutz) within the project ''AutoQML`` and project numbers 01MQ22002A and 01MQ22002B.
\end{acknowledgments}

\section*{Contribution Statement}
\textbf{D.B.}: Conceptualization, Methodology, Software, Validation, Formal analysis, Investigation, Writing, Visualization. \textbf{J.B.}: Conceptualization, Methodology, Software, Validation, Formal analysis, Investigation, Writing, Visualization. \textbf{C.T.}: Funding Acquisition, Supervision.
\textbf{F.S.}: Conceptualization, Supervision, Funding Acquisition.

\section*{Data Availability}
The data used in this study is available upon reasonable request.

\section*{Competing interests}
The authors declare no competing interests.

%

\end{document}